\documentclass[%
preprint,
 amsmath,amssymb,
 aps,
]{revtex4-2}

\usepackage{graphicx}
\usepackage{dcolumn}
\usepackage{bm}

\usepackage{newtxtext}
\usepackage{newtxmath}
\usepackage{natbib}
\usepackage{hyperref}

\usepackage{multirow}
\usepackage{epstopdf, epsfig}
\usepackage[dvipsnames]{xcolor}

\usepackage{tikz,siunitx} 
\usepackage{pgfplots}

\usepackage{longtable}
\usepackage{mwe} 
\usepackage{amsmath, amssymb}
\usepackage{mathrsfs}
\usepackage{placeins}
\usepackage{upgreek}
\usepackage{xkeyval,xcolor}
\usepackage{natbib}

\usepackage{mathtools}

\usepackage{dirtytalk}

\usepackage[utf8]{inputenc} 

\DeclareUnicodeCharacter{00A0}{~}

\hypersetup{
    colorlinks = true,
    urlcolor   = blue,
    citecolor  = black,
}
\newcommand{\RomanNumeralCaps}[1]
\linenumbers

\pgfplotsset{compat=1.18} 
\def\mathclap#1{\text{\hbox to 0pt{\hss$\mathsurround=0pt#1$\hss}}}

\makeatletter
\newlength{\sfp@hseplen}\newlength{\sfp@vseplen}
\define@cmdkey{subfigpos}[sfp@]{pos}[ul]{}
\define@cmdkey{subfigpos}[sfp@]{font}[\small]{}
\define@cmdkey{subfigpos}[sfp@]{vsep}[1\baselineskip]{\setlength{\sfp@vseplen}{\sfp@vsep}}
\define@cmdkey{subfigpos}[sfp@]{hsep}[35pt]{\setlength{\sfp@hseplen}{\sfp@hsep}}
\newcommand{\subfigimg}[3][,]{%
  \setkeys{Gin,subfigpos}{pos,font,vsep,hsep,#1}
  \setbox1=\hbox{\includegraphics{#3}}
  \ifnum\pdfstrcmp{\sfp@pos}{ul}=0
    \leavevmode\rlap{\usebox1}
    \rlap{\hspace*{\sfp@hsep}\raisebox{\dimexpr\ht1-\sfp@vsep}{\sfp@font{#2}}}
    \phantom{\usebox1}
  \else\ifnum\pdfstrcmp{\sfp@pos}{ur}=0
    \leavevmode\usebox1
    \llap{\raisebox{\dimexpr\ht1-\sfp@vsep}{\sfp@font{#2}}\hspace*{\sfp@hsep}}
  \else\ifnum\pdfstrcmp{\sfp@pos}{lr}=0
    \leavevmode\usebox1
    \llap{\raisebox{\sfp@vsep}{\sfp@font{#2}}\hspace*{\sfp@hsep}}
  \else
    \leavevmode\rlap{\usebox1}
    \rlap{\hspace*{\sfp@hseplen}\raisebox{\sfp@vsep}{\sfp@font{#2}}}
    \phantom{\usebox1}
  \fi\fi\fi
}
\makeatother

\begin{document}

\preprint{APS/123-QED}

\title{Experimental evidence of tonal noise and its control in tip-leakage flows}

\author{Prateek Jaiswal}
 \email{jaiswalprateek@protonmail.com}
 \affiliation{Department of Aeronautics and Astronautics, University of Southampton, SO16 7QF, UK.}
 \altaffiliation[Also at ]{
 Department of Mechanical Engineering, University of Sherbrooke, Sherbrooke,  QC J1K 2R1, CA
}%

\author{St{\'e}phane Moreau}%
\affiliation{%
 Department of Mechanical Engineering, University of Sherbrooke, Sherbrooke,  QC J1K 2R1, CA}%




\date{\today}

\begin{abstract}

The present study is first to provide experimental evidence of tonal noise and acoustic feedback by a stationary airfoil with a tip-gap placed between two side plates. Tones are linked to an aero-acoustic coupling using synchronized measurements between Particle Image Velocimetry and far-field microphone probes. To aid in the correlation analysis, the wall boundary condition on the pressure side of the airfoil was modified by introducing surface roughness.

For tip leakage flows without roughness elements, the far-field acoustic spectra show peaks in the acoustic autospectra. These peaks are present regardless of the flow speed studied. The findings reveal that tones associated with tip-gap extend till 16.5 kHz in a low-noise facility. Roughness alters cross-flow velocity along the airfoil chord, influencing acoustics at matched Reynolds and Mach numbers. Roughness induces cancellation of spectral peaks and changes in the velocity and acoustic pressure correlation pattern between airfoil pressure and suction side. Modal shapes confirm the presence of coherent structures in the tip-gap region, with smooth surfaces exhibiting confined flow instabilities and non-dipolar noise sources around and beyond 10 kHz. At these high frequencies, the sound radiation mechanism can be uniquely attributed to neither a jet-like noise source nor these acoustic peaks can be uniquely attributed to coherent structures in the tip-gap region. On the contrary, coherence between airfoil sides is required for noise peaks, which stresses the importance of diffraction on the airfoil pressure side and acoustic feedback.

\end{abstract}

\maketitle


\section{Introduction.}
\label{sec:headings}

One of the main contributors to fan broadband noise is tip noise, which is generated through the interaction between tip vortices, the casing boundary layer, and the blade tip \citep{moreau2018advanced,moreau2019turbomachinery,Moreau2024}. Furthermore, if the tangential velocity of these vortices is smaller than the local entrainment velocity, they may interact periodically with the blade yielding large subharmonic humps, as has been observed in low-speed fans \citep{Moreau:2016}. Our limited understanding of tip-leakage-flow (TLF) noise partly stems from the difficulty of distinguishing individual noise sources in a representative fan rig. To address this issue, \citet{Grilliat2007} proposed a mock-up consisting of a stationary cambered airfoil with a tip gap, to study its noise-generating mechanisms. Although the range of tip gaps studied to date may not be representative of a real turbofan, the mechanisms by which aerodynamic perturbations are converted into sound should be similar \citep{camussi2010experimental}. In this particular set-up, several authors \cite{jacob2016time,saraceno2022tip} have measured high-amplitude peaks and humps in the acoustic spectra. However, the precise reason for the origin of these acoustic peaks is not well understood, and several questions related to them remain unanswered. For instances, at what frequencies do these peaks operate? Can these acoustic tones be uniquely attributed to coherent structures in the tip-gap region? Therefore, the present manuscript seeks to unravel structure of turbulent shear flows that gives rise to these acoustic tones, and to delineate an aero-acoustic coupling between the source and far-field acoustics. 


Tip gap between a rotor and duct in turbomachinery is essential. However, complex vortical structures are generated by the tip-leakage flow and its interactions with
the boundary layer developing over the duct. The tip-leakage vortical structures often induce instabilities and blockage in the flow, which may result in loss in performance \cite{furukawa1998effects}. Furthermore, in an axial compressor, an increase in the tip gap has been shown to lead to an increase in the size of the tip leakage vortex and an increase in the angle between the tip leakage vortex and the airfoil wake \cite{goto1992three}. \citet{muthanna2004wake} studied the effects of the tip gap on the structure of the tip leakage vortex by varying the incoming turbulence and the thickness of the boundary layer. They had varied the size of the tip gap from $0.8\%$ of the airfoil chord to $3.3\%$. They reported an increase in mean velocity, vorticity, and turbulent kinetic energy as the size of the tip gap increased. In summary, tip leakage flows have been the subject of intensive research for the past several decades.

On the other hand, only a handful of studies have studied noise generated by tip-leakage flows. Previous aeroacoustic studies~\cite{camussi2010experimental} showed that at frequencies below 5 kHz, structures within the tip gap region radiate noise as they convect past the trailing edge corner; \citet{camussi2010experimental} achieved this by analyzing temporal changes in averaged pressure signals. On the other hand, numerical simulations performed by \citet{koch2021large}, point out that in this frequency range (1-5 kHz) the dominant contribution is due to the Tip Separation Vortex (TSV). \citet{saraceno2022tip} and \citet{jacob2016time} have noted that tip gap noise operates within the range of $0.7-10$~kHz. Although \citet{camussi2010experimental} attributed the low-frequency mechanism corresponding to a Strouhal number based on the chord length $C$ and the free-stream velocity $U_\infty$, St$_c$ less than 10 to dipole-like noise sources, they were unable to elucidate the sound-radiation mechanism at high frequency. Subsequently, \citet{jacob2010aeroacoustic} hypothesized, based on $U_\infty^7$-$U_\infty^8$ scaling of the acoustic pressure \cite{lighthill1952sound} and detailed flow measurements, that at high frequencies between 4 and 7 kHz (St$_c \ge 10$), jet-like clearance flow may generate sound when leaving the tip-gap region. However, due to high background noise in their facilities, it was not possible to make acoustic measurements beyond $10$ kHz. While no Direct Numerical Simulation (DNS) has yet been performed for this case, Zonal Large-Eddy Simulations (ZLES) by \citet{boudet2016zonal} and Large-Eddy Simulations (LES) by \citet{koch2021large} have achieved good agreement with experiments. The latter also suggested that both the tip leakage vortex and the tip separation vortex contribute to the tip noise generation. However, these simulations rely on the use of acoustic analogies to estimate far-field noise, and with a LES resolution, high-frequency source terms cannot be fully resolved. In summary, limited data are available on the tip gap noise of a single fixed airfoil at high frequency, and little evidence on a jet-like noise mechanism has been reported \cite{Grilliat2007,jacob2010aeroacoustic}; therefore, the noise generated by a fixed airfoil with tip-gap needs further exploration.

More recent experiments \cite{jacob2016time,moreau2016wall,saraceno2022tip} have reported the presence of high-amplitude peaks and humps in the acoustic spectra. The peak in acoustic spectra reported by \citet{jacob2016time} appears to be localized in a narrow band of frequencies, which is reminiscent of the acoustic feedback observed in flows past the cavity \cite{rossiter1964wind}, and in the presence of a Kelvin-Helmoltz type instability encountered in the flow transition over airfoils \cite{Sanjose2019,jaiswal2022experimental}. More importantly, the framework used in acoustic analogies, often used in numerical simulations to estimate far-field noise \cite{boudet2016zonal,koch2021large}, does not permit any feedback loop or coupling between acoustic and aerodynamics~\cite{glegg2017aeroacoustics}, which can lead to frequency selection \cite{jaiswal2022experimental}. Therefore, it is unclear whether tip noise operates only within the $0.7-10$~kHz range or if the origins of these tones are purely aerodynamic in nature?

Therefore, the objective of the present study is two-fold:

1) Explore the range in which the tip noise operates. 

2) Provide experimental evidence on the genesis of humps or tones in the acoustic spectra.

To this end, aeroacoustic measurements were performed in the anechoic wind tunnel at Universit{\'e} de Sherbrooke (UdeS)~\cite{jaiswal2022experimental}. In particular, planar Particle Image Velocimetry (PIV) measurements in the tip gap and acoustic pressure measurements at a far-field location were performed in a synchronized manner. Lastly, previous studies \cite{Grilliat2007,camussi2010experimental,jacob2016time,saraceno2022tip} have mostly been conducted at a high angle of attack, which does not reflect the actual TLF induced by a fan operating at its design condition. Therefore, in the present study measurements are performed at a fixed geometric angle of attack of $8^{\circ}$ for the controlled diffusion (CD) airfoil, which corresponds to an on-design flow conditions for this specific airfoil geometry used in multiple turbomachinery applications (both low-speed and high-speed fans and compressors)~\cite{Roger2004,Moreau2005,Wang2009,neal2010effects,lallier2013numerical,Moreau:2016,Wu2020,jaiswal2020use}. The paper is structured as follows. Details on the experimental setup, methodology can be found in section \ref{sec:1}. Section \ref{sec_results} reports the experimental findings. Section \ref{sec:disc} provides a discussion on key findings. Finally, conclusions and perspectives are drawn in section~\ref{sec:conclusion}.

\section{Experimental Set-up and Instrumentation:} \label{sec:1}

All measurements were performed in the anechoic wind tunnel at UdeS at a free-stream velocity $U_\infty$ of 16~m/s and 28~m/s, which corresponds to Reynolds numbers based on the airfoil chord length $C$ of $Re_c \simeq 1.5\times 10^5$ and $Re_c \simeq 2.5\times 10^5$ respectively. The anechoic wind tunnel consists of an anechoic room, which is approximately $7 \times 5.5 \times 4$ m$^3$ in dimension and is coupled with an open jet with a nozzle size of $50 \times 30$ cm$^2$. The mock-up of the CD airfoil used in the present study has a 0.1347 m chord, and a 0.3 m span, with a $12^{\circ}$ camber angle. It has  4$\%$ thickness-to-chord ratio. The thin airfoil is placed at a $8^{\circ}$ geometric angle of attack with the help of plexiglass plates of thickness 4.25~mm laser cut to reduce uncertainty in angle of attack while placing the airfoil and at the same time giving good optical access. The maximum turbulence intensity of the facility is less than $0.4\%$ \cite{Moreau2016}. As such, the resulting laminar boundary thickness that develops over the side plate was estimated to be between 1.9 and 1.4 mm for the 16 and 28 m/s cases respectively, based on the Blasius boundary layer \cite{schlichting2016boundary}. These estimates are consistent with the thickness of the incoming boundary layer measured in a similar open-jet facility with  a single hot wire by~\citet{neal2010effects}. The thickness of the boundary layer close to the trailing edge on the airfoil suction side was estimated to be $6.34$ mm~\cite{jaiswal2020use}, for the CD airfoil placed at $8^{\circ}$ geometrical angle-of-attack and 16 m/s.

\begin{figure}
\centering
  \includegraphics[width=0.8\textwidth]{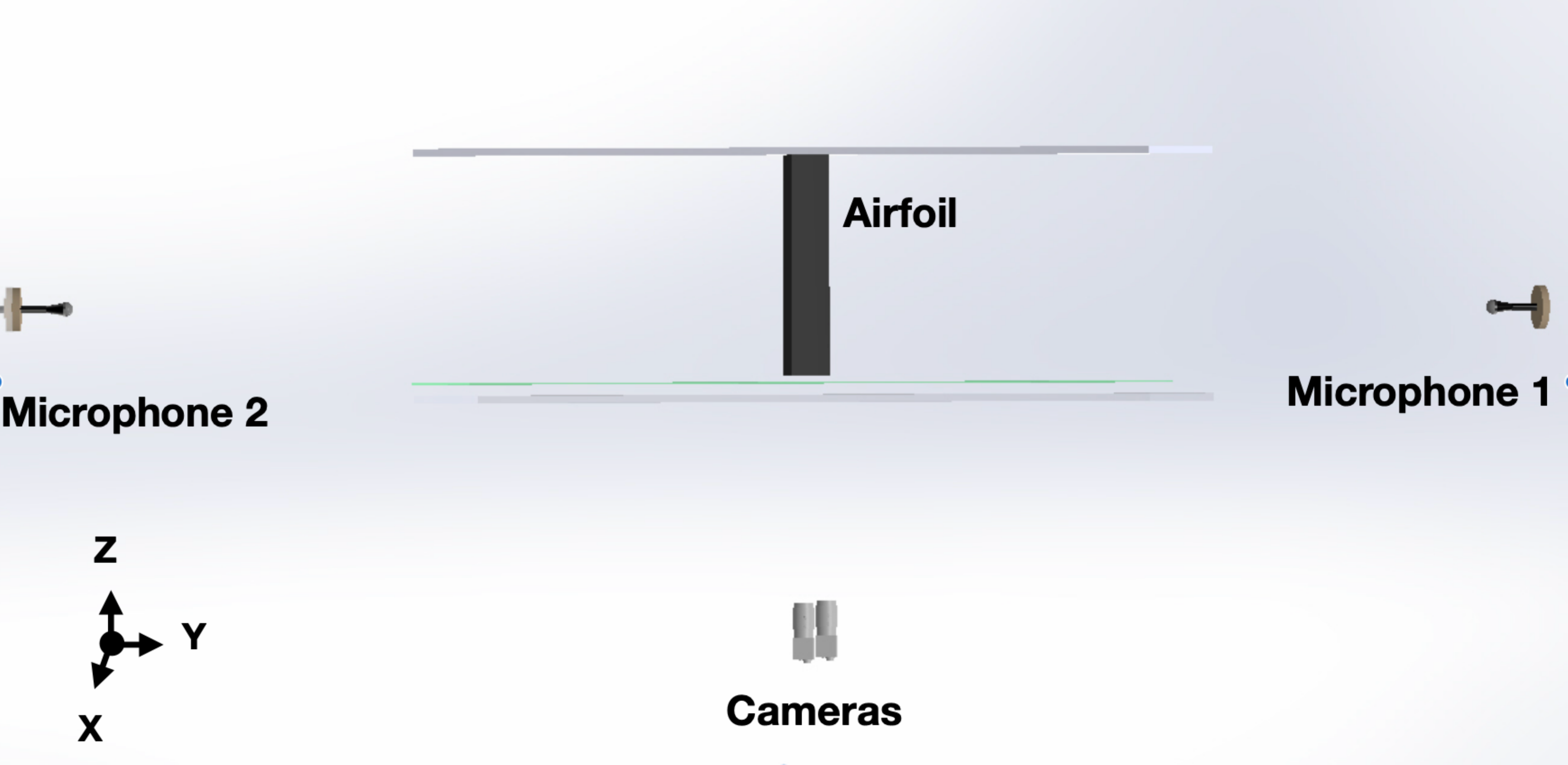}
\caption{Experimental setup. Figure not drawn to scale. Laser sheet is depicted in green. Mean flow is aligned with the X-axis.}
\label{set}       
\end{figure}

An 8 mm tip gap, $t_p$, which corresponds to approximately 2.5\% of airfoil span, 6\% of airfoil chord length, and 150\% of the maximum airfoil thickness, is created between the airfoil and the bottom plate, as shown in Figure~\ref{set}. Such tip parameters are typical of low speed fan systems for which significant tip noise has been reported~\cite{Moreau:2016,Zhu2018,Kroemer2019}. More importantly, the size of the incoming boundary layer, estimated to be less than 2~mm, is smaller than the size of the tip gap. Recently, \citet{saraceno2023influence} have shown that the ratio between the tip-gap size and the boundary layer has a minimal effect on tip-leakage noise, especially when this ratio is greater than 1. Furthermore, in the present study, the size of the tip gap (relative to airfoil chord length) 
also corresponds to one of the three cases studied by \citet{you2006effects}, who showed that the fundamental features of tip leakage flows remain unaltered within the range of tip-gap size investigated in their study. As such, the present study is relevant for applications where the noise generated by the tip leakage flow is a source of concern. Finally, additional tests were performed by covering about 10\% of the airfoil span with a P36 grit roughness (grain size $\approx 525$ $\mu$m) on the pressure side of the airfoil. The strip of roughness was placed between $0.1-0.8\, C$ at the tip of the airfoil.

\subsection{Planar PIV measurements setup}

Two-dimensional planar PIV measurements were performed in the tip gap region, about $2$ mm below the airfoil, as shown in Figure~\ref{set}. Two sCMOS cameras, each with $5.5$ megapixel sensors, were used to acquire images in dual frame mode. An ND:YAG dual pulsed laser was used for illumination. A thin ($\le 0.5$ mm) light sheet was generated with a set of spherical lenses and a divergent cylindrical lens. Tracer particles of about 1~$\mu$m 
were generated to seed the flow. In total, more than $1800$ images were recorded for each case at an acquisition frequency of 4~Hz. The maximum mean particle image displacement was more than 20 pixels, which ensures the relative error is within  $\sim 0.5 \% $ for the estimation of particle image displacement. The data collected were processed using Lavision's Davis 8. For vector calculations, an iterative multigrid scheme was used, with a final window size of $16\times 16$ pixels and an overlap of 75~\%. For the first camera, placed in the leading edge region, the final window size was about $0.28 \times 0.28$ mm$^2$. In comparison, for the second camera, which was placed in the trailing edge region, the final window size was approximately $0.23 \times 0.23$ mm$^2$.

\subsection{Acoustic measurements}

Far-field acoustic pressure measurements were performed with two $1/2$ inch Integrated Circuit Piezoelectric (ICP) microphones, placed on the suction and pressure sides of the airfoil. These microphones were placed perpendicular to the trailing edge of the  airfoil and placed at a distance ($R$) of $\sim 1.34$~m, which is about 10 times the airfoil chord length, to ensure that they are in an acoustic far-field location. The microphones were calibrated using a B\&K piston-phone, which ensures that the calibration uncertainty is within 0.2~dB. For noise measurements, a long time signal (3 min) was used, which yielded an uncertainty of $\approx 0.3$ dB ($95\%$ confidence) in the estimation of acoustic spectra \citep{bendat1978statistical}. Finally, note that the 90$^\circ$ position of the microphones with respect to the airfoil chord is selected to maximize the expected dipolar noise radiation from the trailing edge~\cite{Roger2005,Moreau2009}, and to minimize the diffraction effects from the wind-tunnel nozzle~\cite{Moreau2007,Moreau2009}. \citet{koch2021large} also showed that this is also the direction in which tip noise is expected to be more prominent (Figure~21~(b) in \citet{koch2021large}).

\subsection{Synchronized 
measurements}

The PIV and far-field acoustic measurements mentioned above were performed in a synchronized manner. While the far-field pressure measurements are time resolved, PIV measurements have a limited time resolution. As such, to synchronize these two measurements, the acquisition frequency for all the measurements performed are set to powers of two. In particular, the PIV measurements were performed at 4~Hz while the far-field pressure were recorded at an acquisition frequency of 65536~Hz (or $2^{16}$~Hz). The signal from laser is used as a trigger to initiate the measurements, and the methodology for these synchronized measurements follows the seminal work of~\citet{henning2008investigation}. As such, the readers should refer to \cite{henning2008investigation} for more details.

\section{Results} \label{sec_results}

\begin{figure*}
  \centering
  \begin{tabular}{@{}p{0.5\linewidth}@{\quad}p{0.5\linewidth}@{}}
    \subfigimg[width=70 mm,pos=ur,vsep=2pt,hsep=38pt]{(a)}{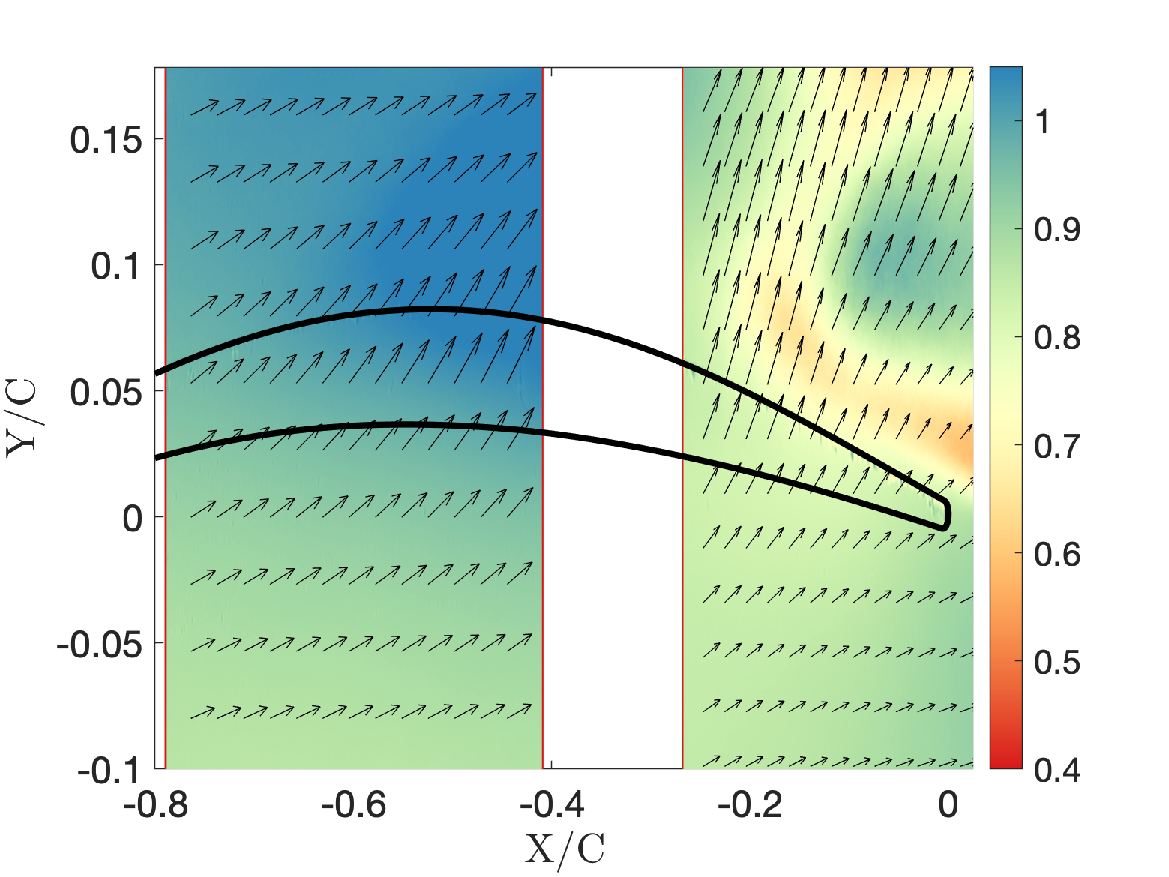} &
    \subfigimg[width=70 mm,pos=ur,vsep=2pt,hsep=38pt]{(b)}{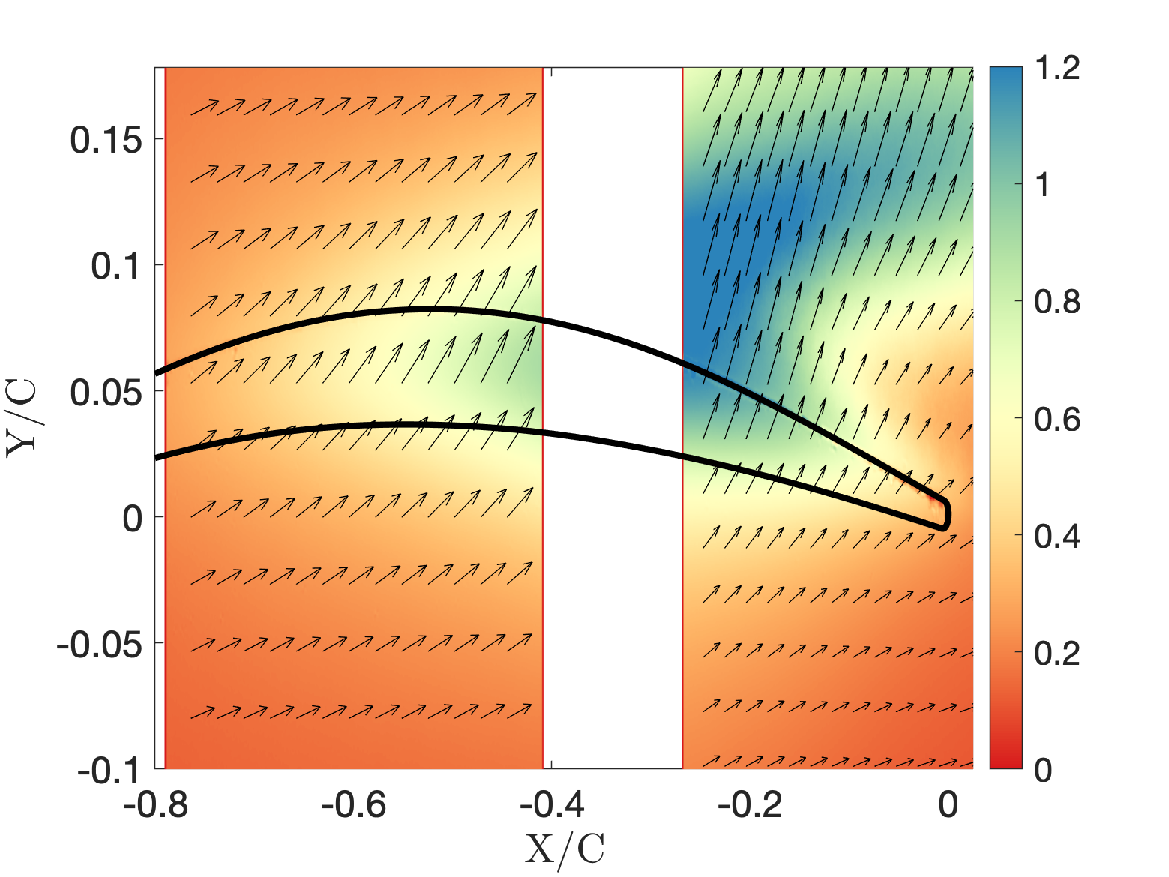} \\
     \subfigimg[width=70 mm,pos=ur,vsep=2pt,hsep=38pt]{(c)}{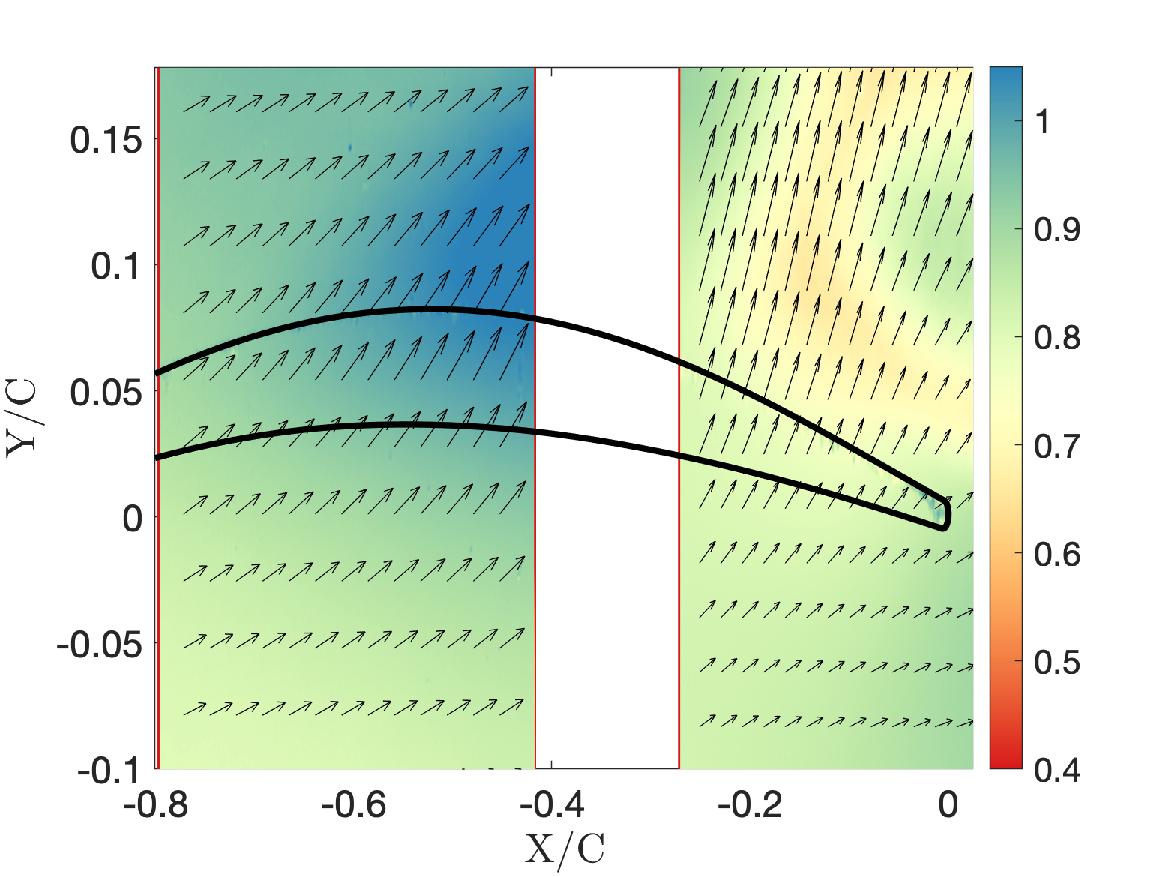} &
    \subfigimg[width=70 mm,pos=ur,vsep=2pt,hsep=38pt]{(d)}{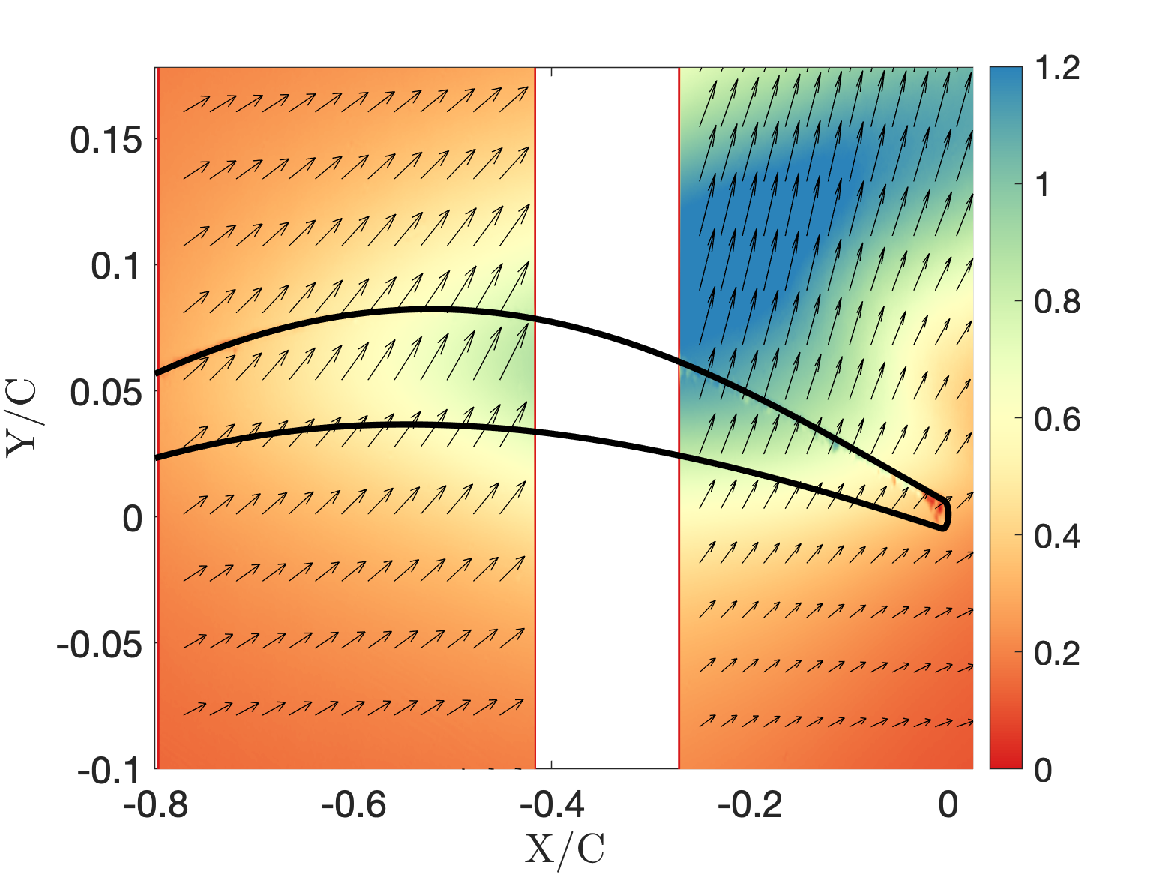} \\
     \subfigimg[width=70 mm,pos=ur,vsep=2pt,hsep=38pt]{(e)}{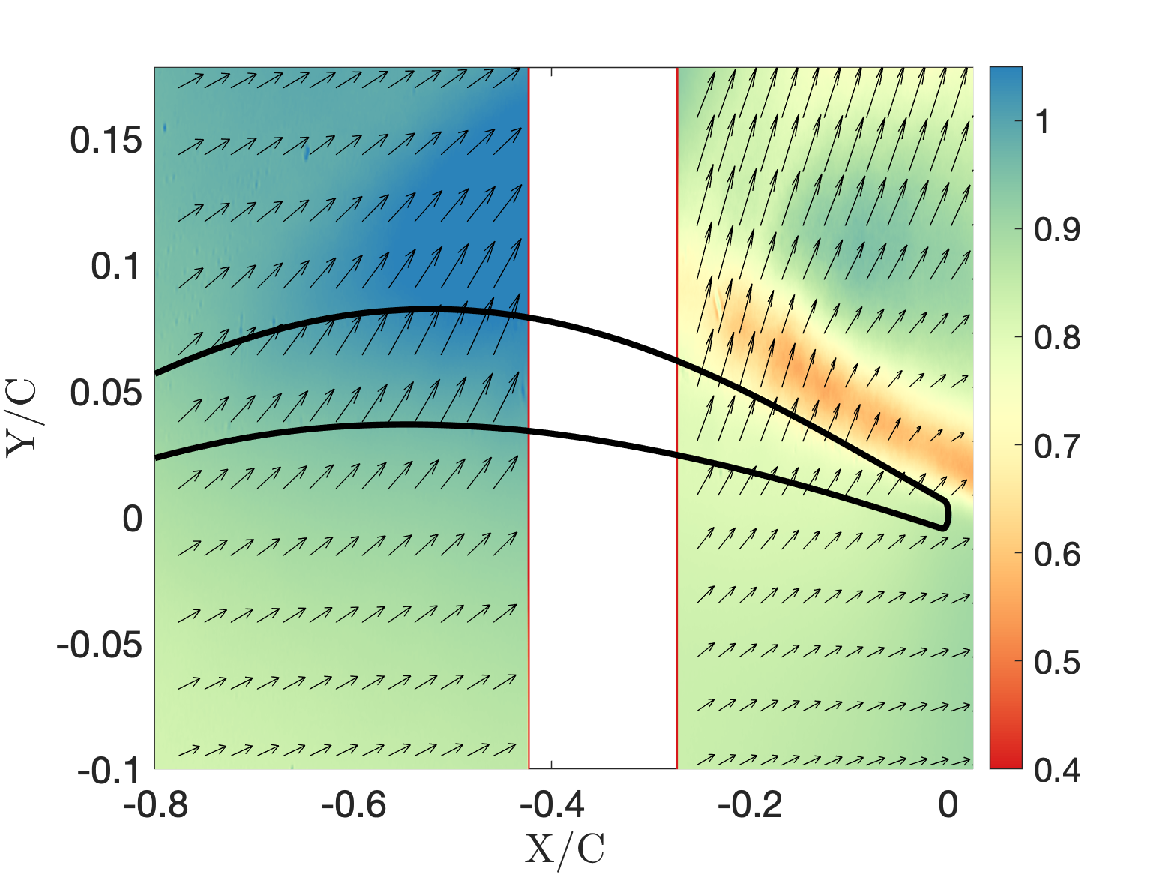} &
    \subfigimg[width=70 mm,pos=ur,vsep=2pt,hsep=38pt]{(f)}{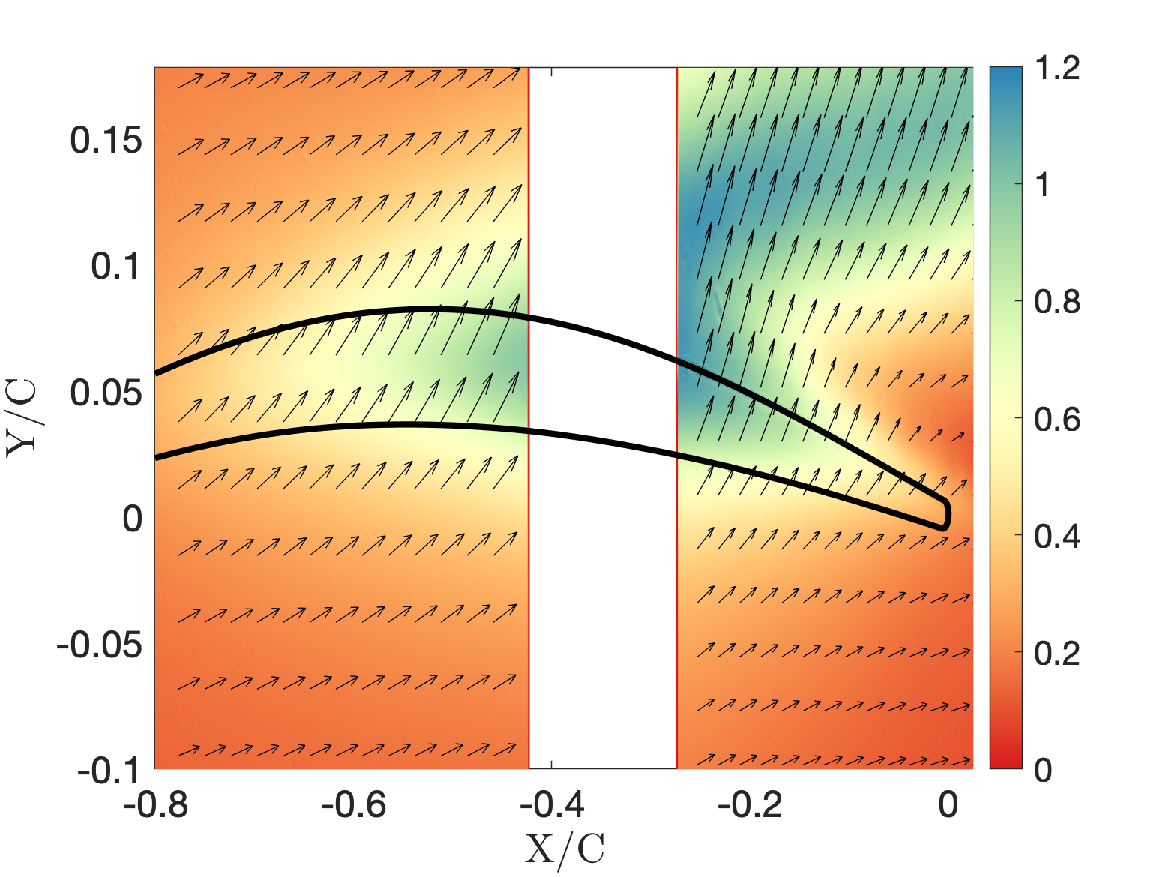} \\
  \end{tabular}
  \caption{Contours of mean velocity over the airfoil tip. Left figures are for $U/U{_{\infty}}$ while right figures for $V/U{_{\infty}}$ Legends: (a) Smooth, 16 m/s (b) Smooth, 16 m/s (c) Smooth, 28 m/s (d)Smooth, 28 m/s (e) Rough, 28 m/s (f) Rough, 28 m/s. }
  \label{fig:mean}
\end{figure*}

Figure \ref{fig:mean} shows the mean velocity in the coordinate system aligned with the wind tunnel for three test cases; 16 m/s (smooth), 28 m/s (smooth), 28 m/s (rough). No measurements were possible between $X/C = 0.3-0.4$ due to 
optical access constraints. The velocity fields have been normalized by the incoming free-stream velocity $U_{\infty}$ in order to compare the velocity maps. The velocity maps for the cases with no roughness (Figures \ref{fig:mean} (a)-(d)) yield similar velocity fields and are comparable despite differences in Reynolds numbers. In contrast, the velocity fields are visibly different in the trailing-edge region for the rough case performed at 28 m/s. In particular, a significant reduction in cross-stream velocity $V/U_{\infty}$ (Figure \ref{fig:mean} (f) versus (d)) can be seen. Furthermore, in the rough case, the cross flow velocity ($V/U_{\infty}$) shows a marginal increase close to the mid-chord location (Figure \ref{fig:mean} (f)), while there is a significant increase in the streamwise direction (Figure \ref{fig:mean} (e)). 

\begin{figure*}
  \centering
  \begin{tabular}{@{}p{0.5\linewidth}@{\quad}p{0.5\linewidth}@{}}
    \subfigimg[width=70 mm,pos=ur,vsep=2pt,hsep=38pt]{(a)}{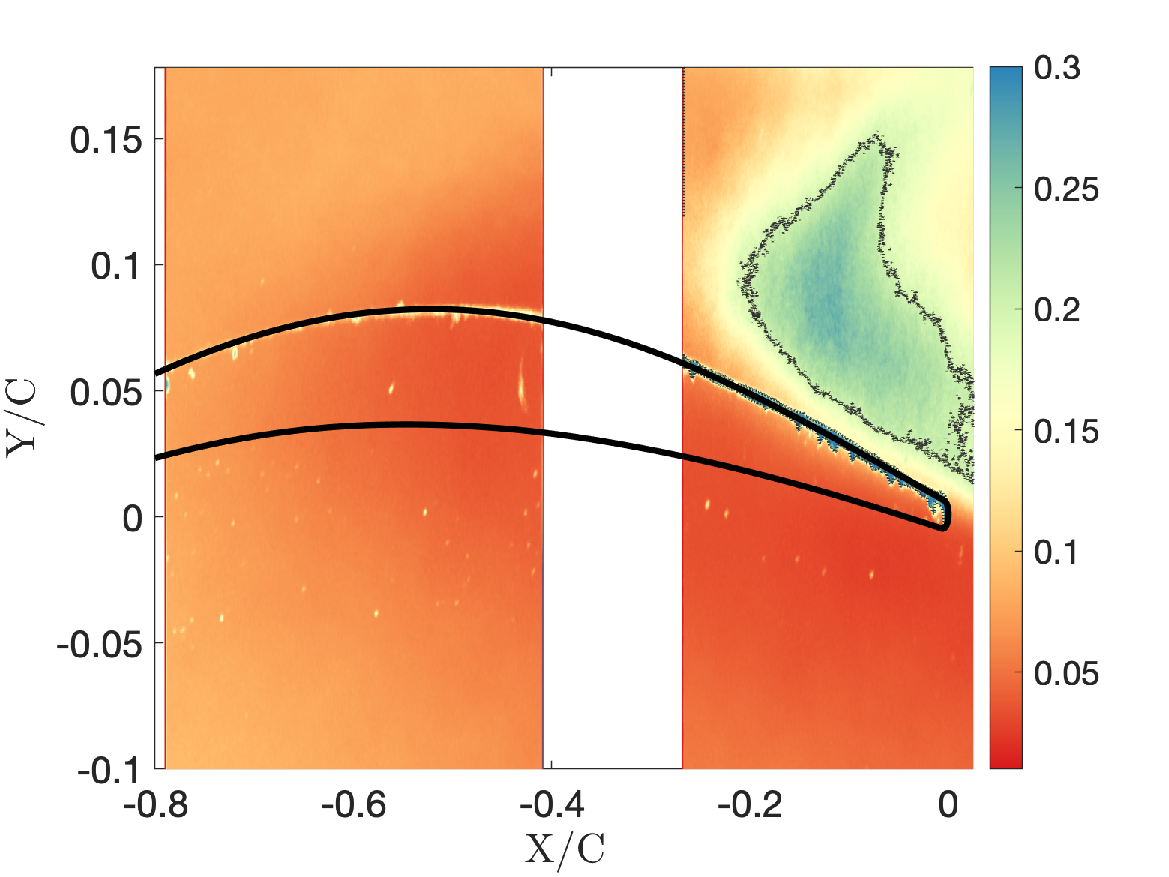} &
    \subfigimg[width=70 mm,pos=ur,vsep=2pt,hsep=38pt]{(b)}{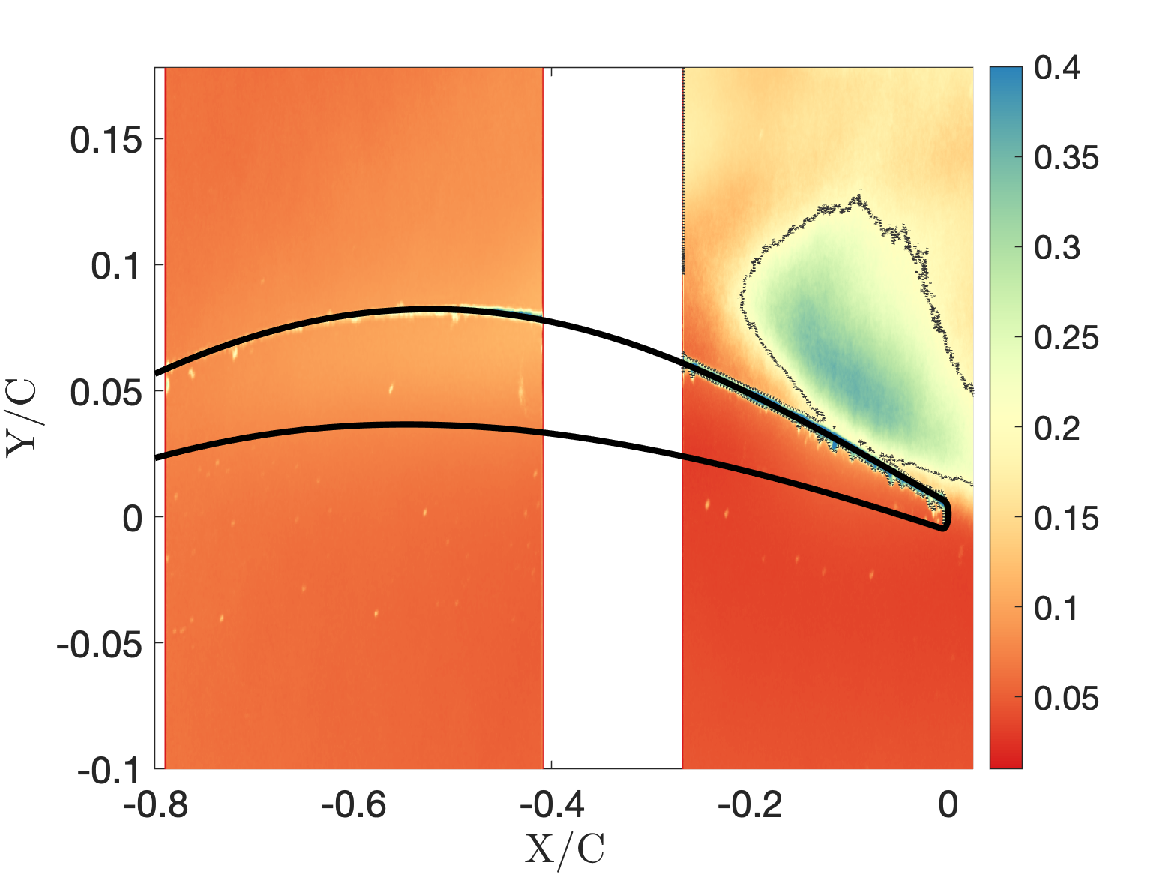} \\
     \subfigimg[width=70 mm,pos=ur,vsep=2pt,hsep=38pt]{(c)}{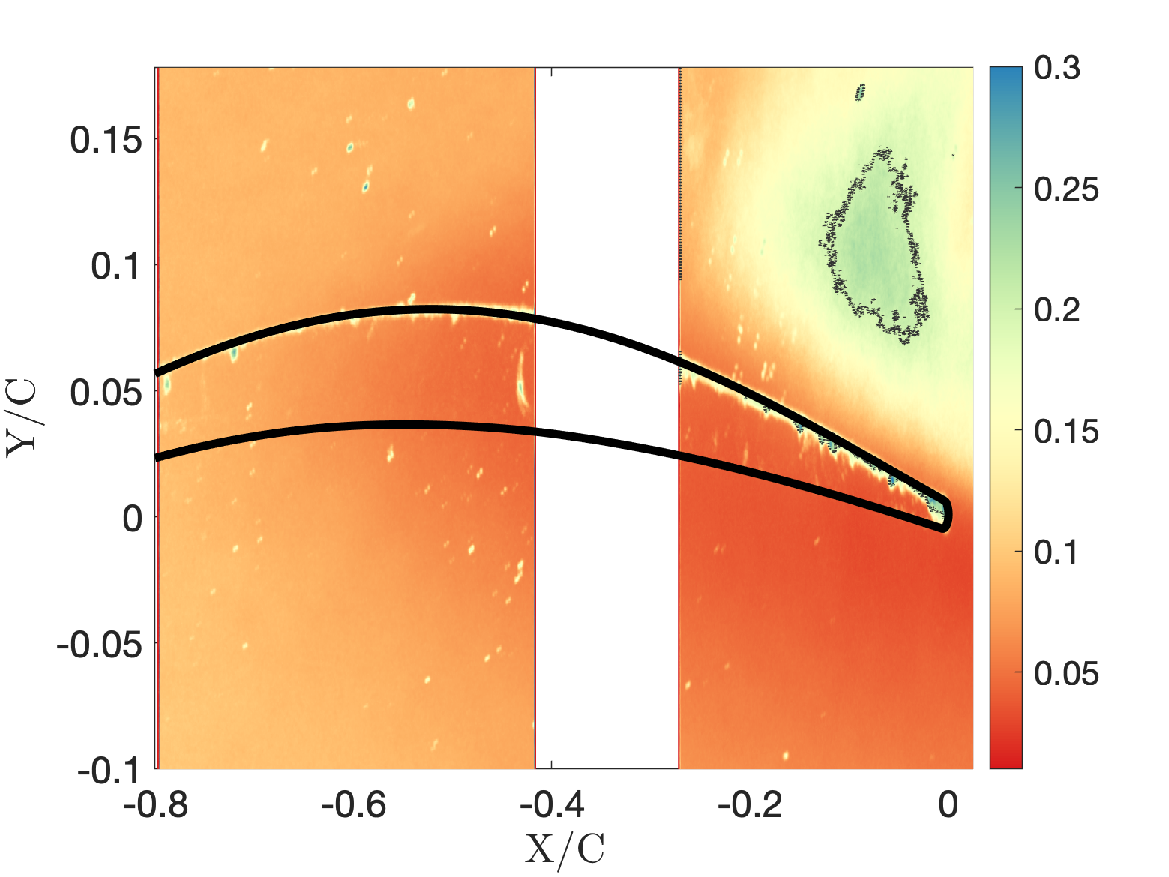} &
    \subfigimg[width=70 mm,pos=ur,vsep=2pt,hsep=38pt]{(d)}{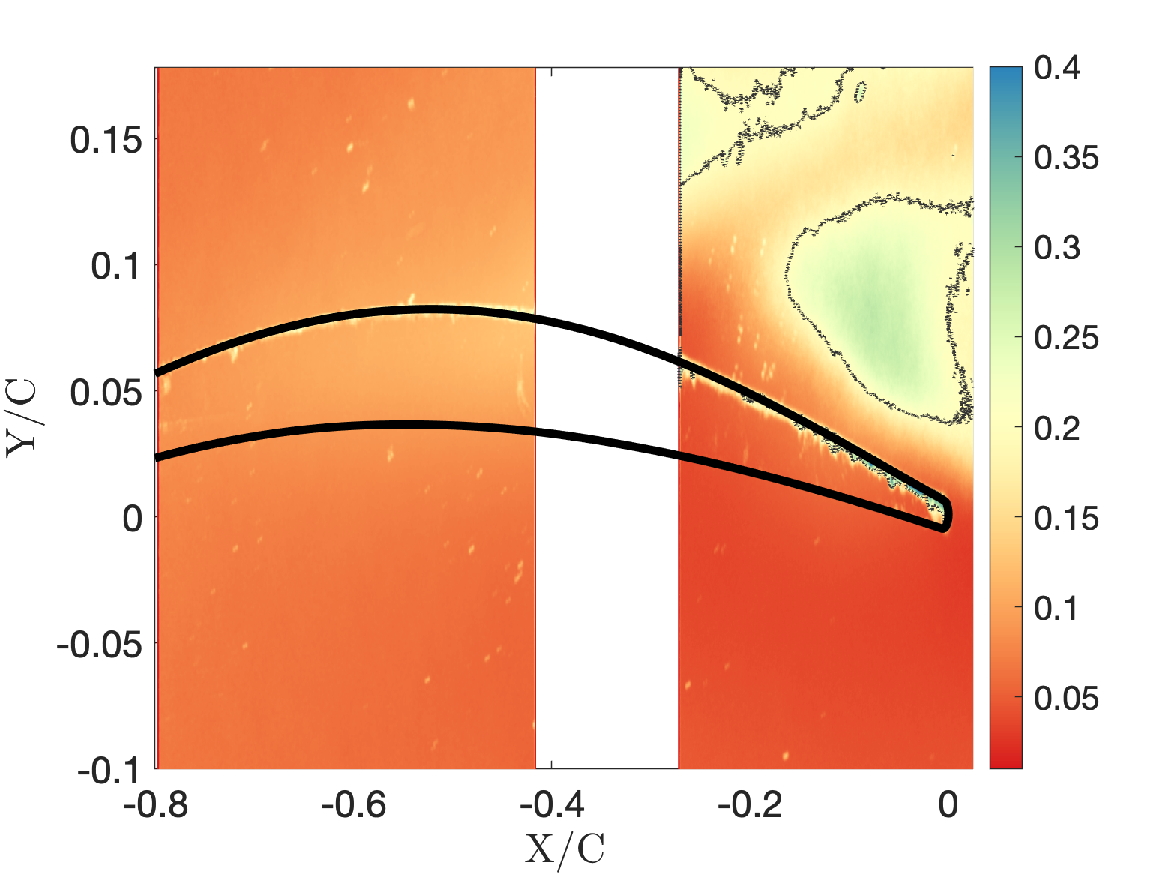} \\
     \subfigimg[width=70 mm,pos=ur,vsep=2pt,hsep=38pt]{(e)}{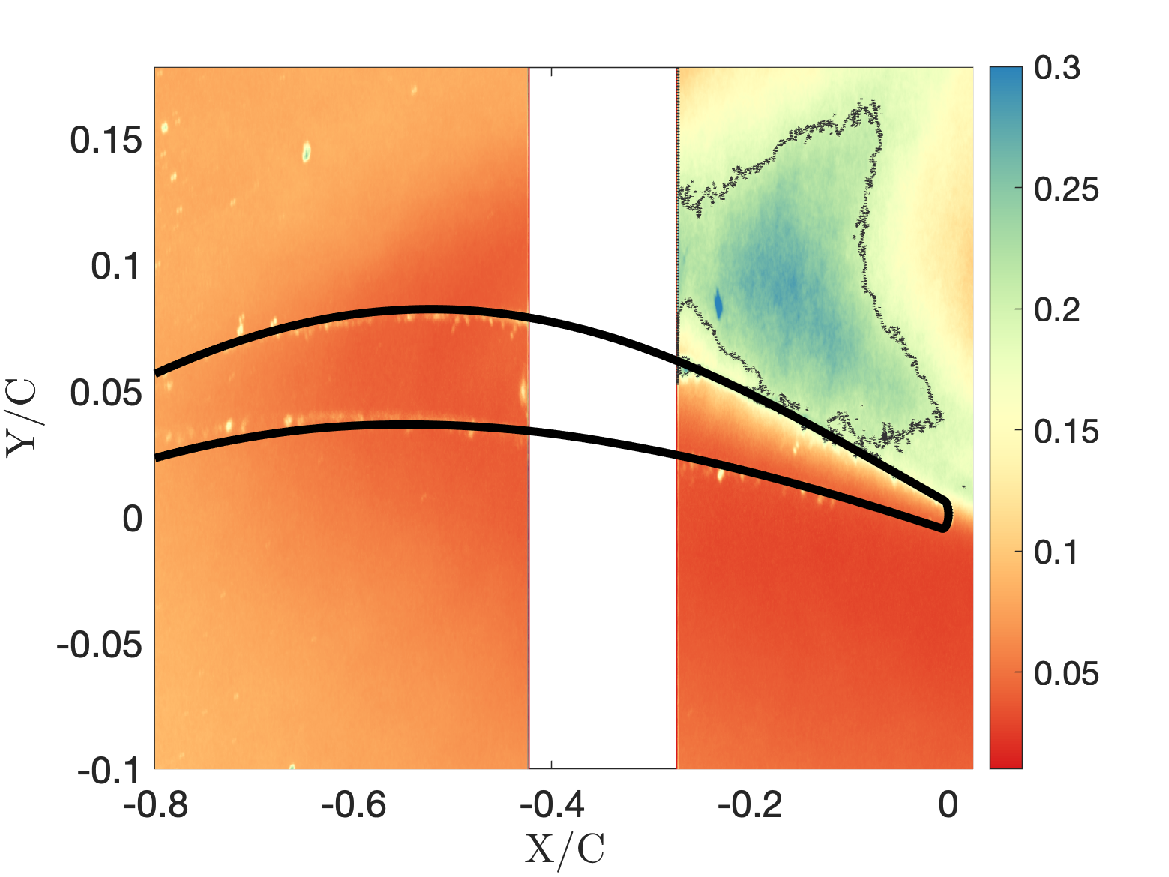} &
    \subfigimg[width=70 mm,pos=ur,vsep=2pt,hsep=38pt]{(f)}{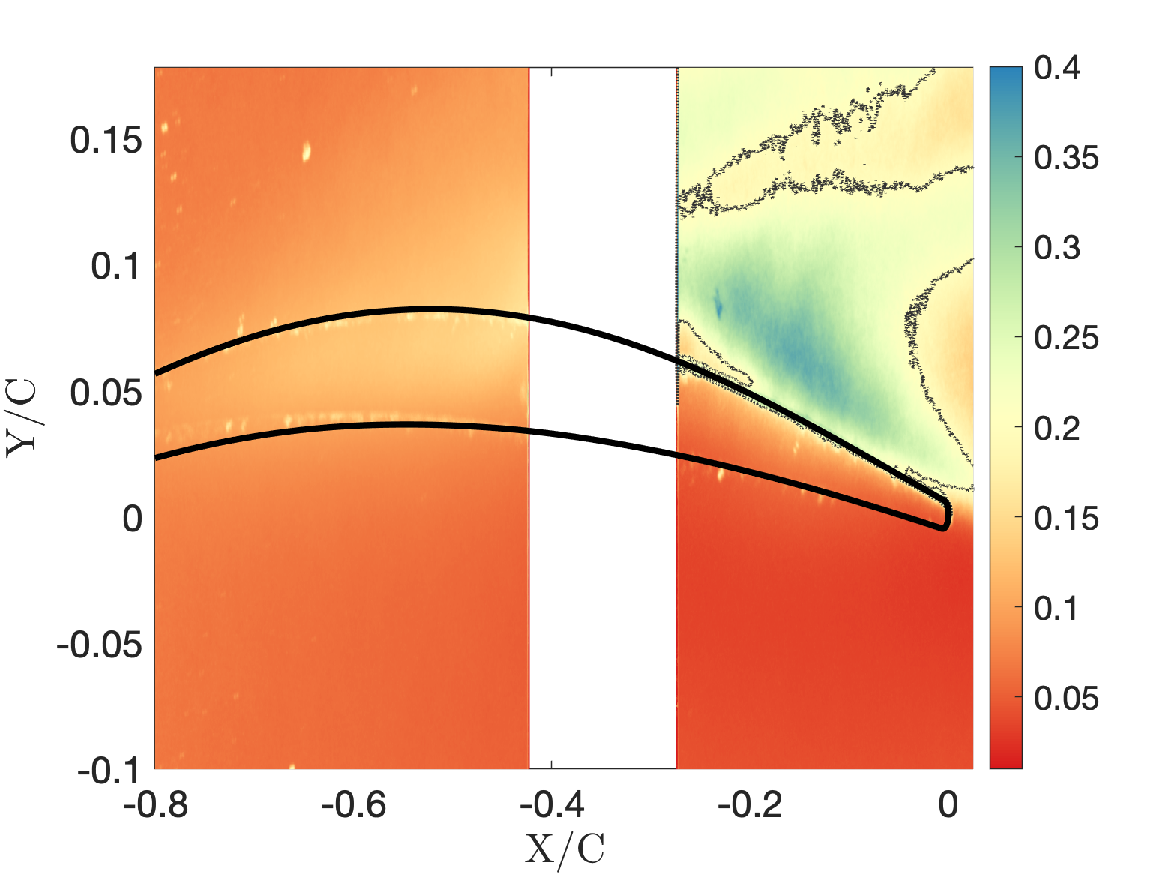} \\
  \end{tabular}
  \caption{Contours of the standard deviation of velocity fluctuations (m/s) measured in the airfoil tip gap region. Left figures are for $U^{\prime}/U{_{\infty}}$ while right figures are for $V^{\prime}/U{_{\infty}}$. Legends: The same as in Figure \ref{fig:mean}. }
  \label{fig:stats}
\end{figure*}

Figure \ref{fig:stats} shows the standard deviation velocity fluctuations 
normalized by the incoming mean free-stream velocity $U_{\infty}$ for the three test cases. The contours of the standard deviation of the velocity fluctuations suggest that the flow remains largely laminar, near the leading edge region, which is in agreement with previous studies \citep{Grilliat2007,saraceno2022tip}. Despite a significant reduction in cross-stream velocity $V/U_{\infty}$ (Figure \ref{fig:mean} (f)) the standard deviation velocity fluctuations $U^{\prime}/U_{\infty}$ and $V^{\prime}/U_{\infty}$ are visible closer to the trailing edge of the airfoil for the rough case. 
Although the overall magnitude of the standard deviation velocity fluctuations is comparable between the three test cases, the latter are localized differently. For both components, the regions of high fluctuations move further downstream and away from the wall with increasing Reynolds number (Figures \ref{fig:stats} (a) and (b) versus (c) and (d)). The clear difference between the rough and smooth cases is that for the latter the regions of elevated velocity disturbance are smaller, more localized compared to the rough wall case. Furthermore, the disturbance field for the rough wall case extends further upstream and closer to the wall (Figures \ref{fig:stats} (e) and (f)).  

\begin{figure*}
  \centering
  \begin{tabular}{@{}p{0.45\linewidth}@{\quad}p{0.45\linewidth}@{}}
    \subfigimg[width=70 mm,pos=ll,vsep=25pt,hsep=22pt]{(a)}{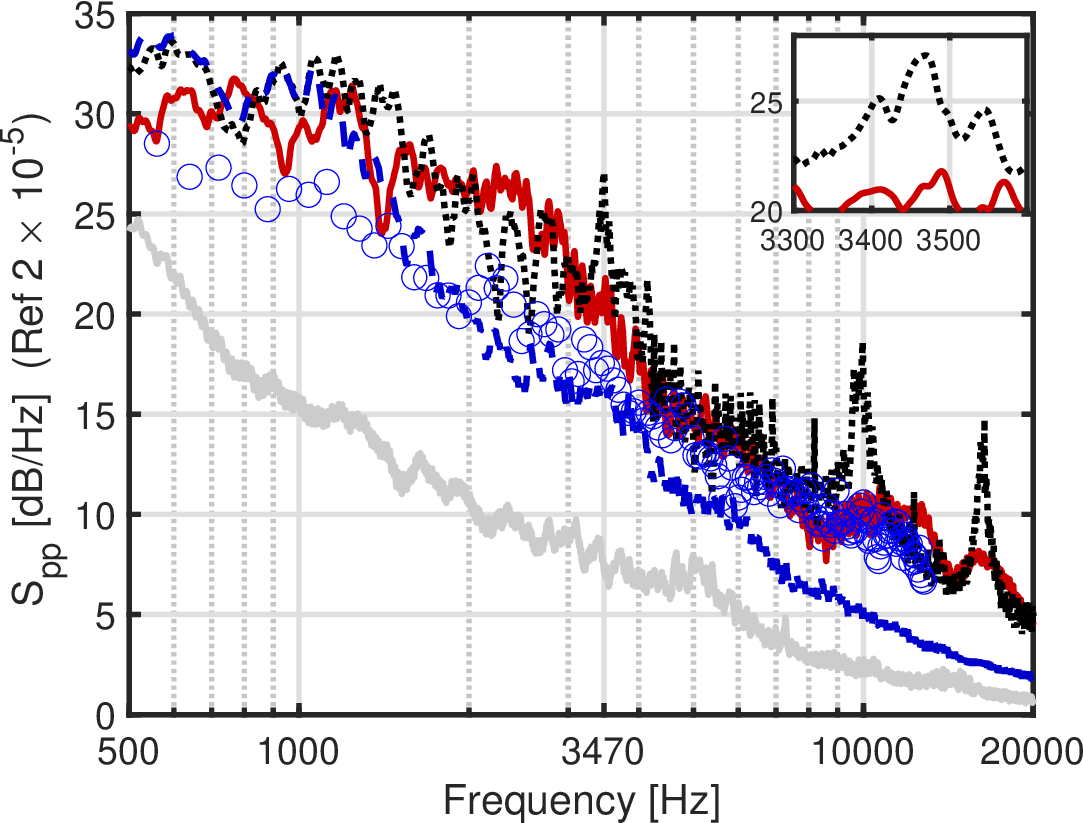} &
    \subfigimg[width=70 mm,pos=ll,vsep=25pt,hsep=22pt]{(b)}{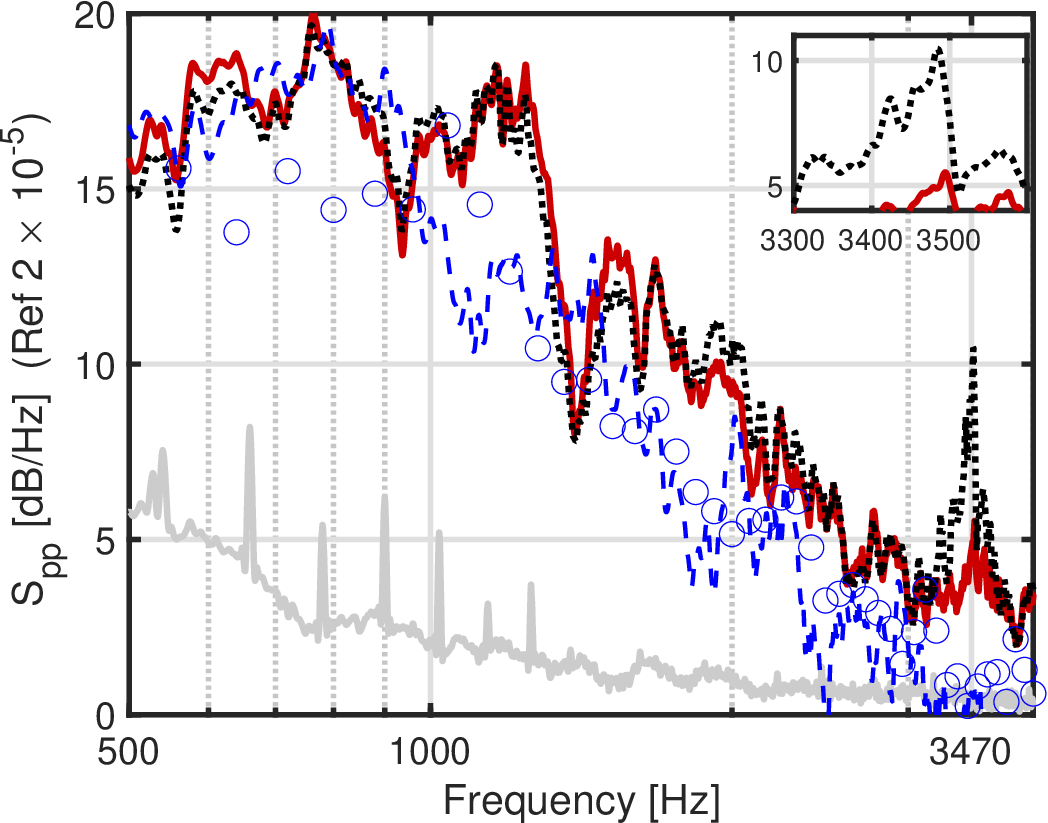} \\
  \end{tabular}

\caption{Sound pressure level measured by an observer at a distance, $R$, of 1.35 m and perpendicular to airfoil chord on its pressure side.
Legends: (a) $U_{\infty}=28$ m/s; (b) $U_{\infty}=16$ m/s. Dotted black line: airfoil with tip gap no roughness; Solid red line: airfoil with tip gap and roughness; Dashed blue line: smooth airfoil with no tip gap; Blue circles: smooth airfoil with no tip gap (ECL facility \cite{Moreau2005}); Dotted gray line: background.}%
\label{Acoustic_spectra} 
\end{figure*}

The Sound Pressure Level (SPL), $S_{pp}$, measured by the microphone placed on the pressure side of the airfoil is shown in figure \ref{Acoustic_spectra}. Note that the SPL measurements by the microphone placed on the suction side were identical; as such, they are not plotted here for clarity. 
SPL was calculated from the time series using Welch's method. In order to accurately estimate spectral properties, a Kaiser window was used with a 75$\%$ overlap. The window size was restricted to $f_s/4$, where $f_s$ is the sampling frequency. Furthermore, zero padding was performed in order to obtain a spectral resolution of 1~Hz. All measurements with the airfoil are clearly above the background jet noise (dotted grey line). The dashed blue line in Figure \ref{Acoustic_spectra} corresponds to the noise generated by an airfoil without a tip gap. As mentioned above, the intensity of the incoming turbulence is $\le 0.4 \%$; therefore, the dashed blue line shows the contribution of the trailing edge noise \cite{brooks1981trailing}. The present trailing-edge noise measurements are also consistent with previous data collected at Ecole Centrale de Lyon (ECL)~\cite{Moreau2005}. Furthermore, the SPL show that the noise generated by the introduction of airfoil tip is dominant above 1~kHz. More importantly, for the first time SPL measurements up to 20~kHz have been achieved due to the UdeS low-noise facility. A tonal peak at 3500~Hz is observed in the SPL, for the cases without roughness elements, as previously reported by \citet{jacob2016time}, who did these measurement for similar tip gap size but at very different Mach and Reynolds numbers. For $U_{\infty}=28$ m/s, two additional peaks, at 10045 and 16387~Hz can be seen for the first time, which also disappear with roughness.


\begin{figure*}
  \centering
  \begin{tabular}{@{}p{0.45\linewidth}@{\quad}p{0.45\linewidth}@{}}
    \subfigimg[width=70 mm,pos=ll,vsep=25pt,hsep=22pt]{(a)}{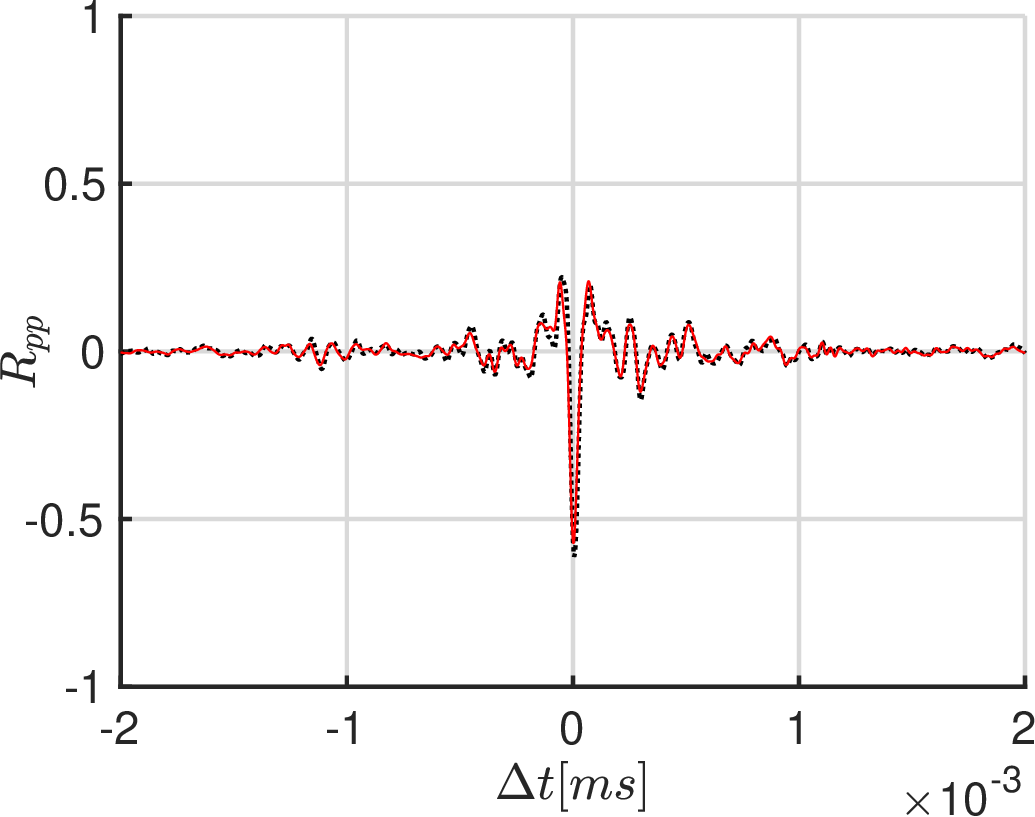} &
    \subfigimg[width=70 mm,pos=ll,vsep=25pt,hsep=22pt]{(b)}{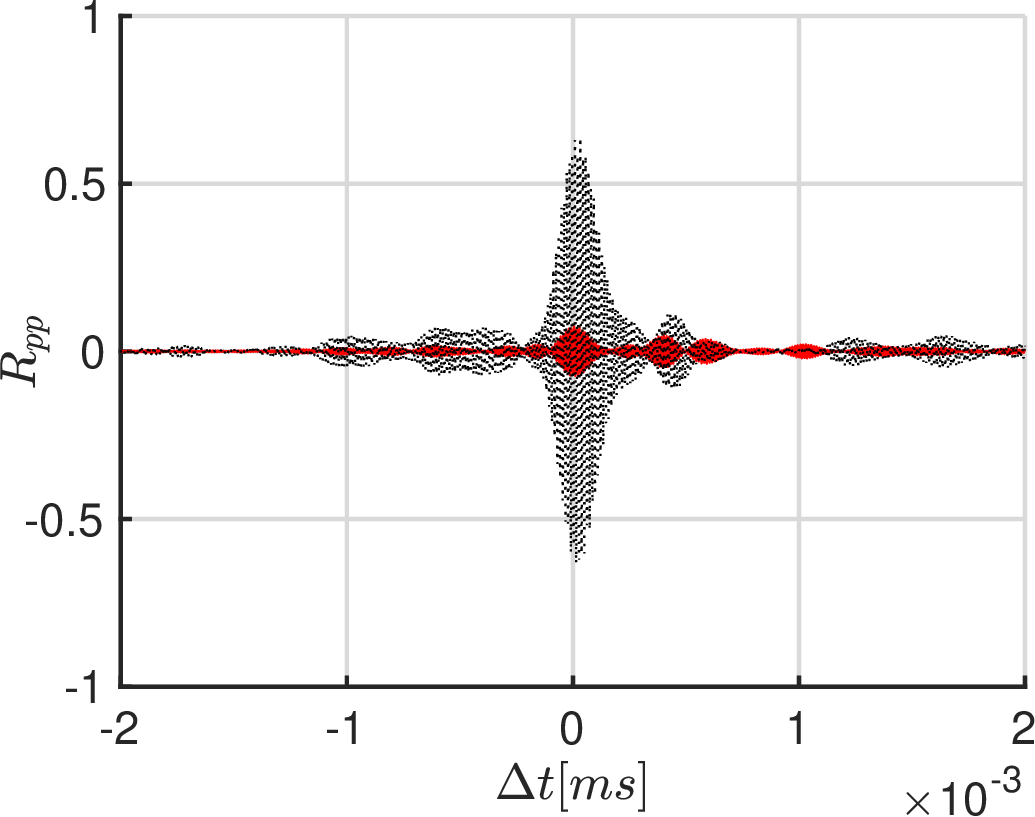} \\
  \end{tabular}

\caption{Far-field sound pressure correlation at $U_{\infty}=28$ m/s. Legends: (a) Band-passed correlation between 500 and 4000~Hz; (b) Band-passed correlation between 9000 and 11000~Hz. Legends: The same as in Figure \ref{Acoustic_spectra}.} 
\label{xcorr} 
\end{figure*}

Figure \ref{xcorr} shows the band-passed correlation $R_{pp}$ between the two opposing far-field microphones for two different frequency bands. In the low frequency band (1000 to 4000~Hz) a clear phase opposition for all cases (negative $R_{pp}$) decisively demonstrates that the dominant noise source is dipolar in nature \citep{yu1978experimental}, consistently with the directivity plots reported by~\citet{koch2021large}. However, at the higher frequency band (9000 and 11000~Hz), the pattern in correlation cannot be explained by dipolar noise sources. Indeed, \citet{jacob2010aeroacoustic} argued that at high-frequencies, the sound pressure levels emitted by the TLF scales with $U_{\infty}^8$, which is typically associated with quadropolar noise sources.

\begin{figure*}
  \centering
  \begin{tabular}{@{}p{0.45\linewidth}@{\quad}p{0.45\linewidth}@{}}
    \subfigimg[width=70 mm,pos=ll,vsep=35pt,hsep=42pt]{(a)}{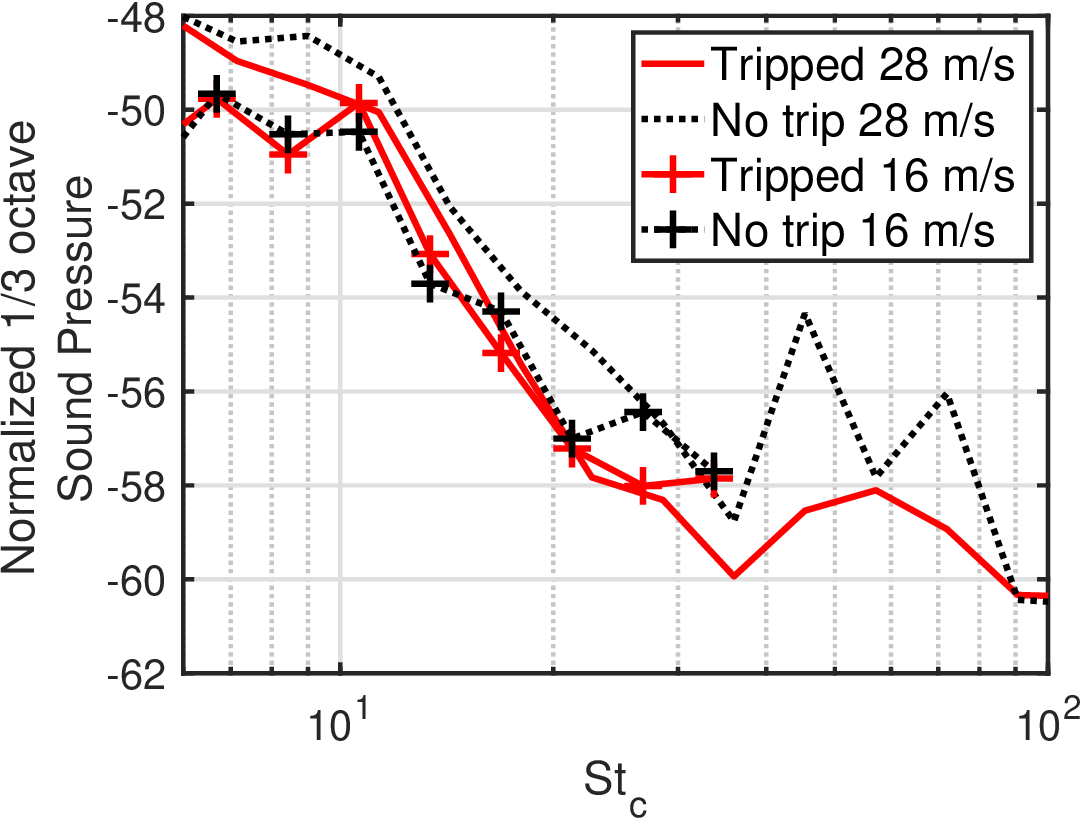} &
    \subfigimg[width=70 mm,pos=ll,vsep=35pt,hsep=42pt]{(b)}{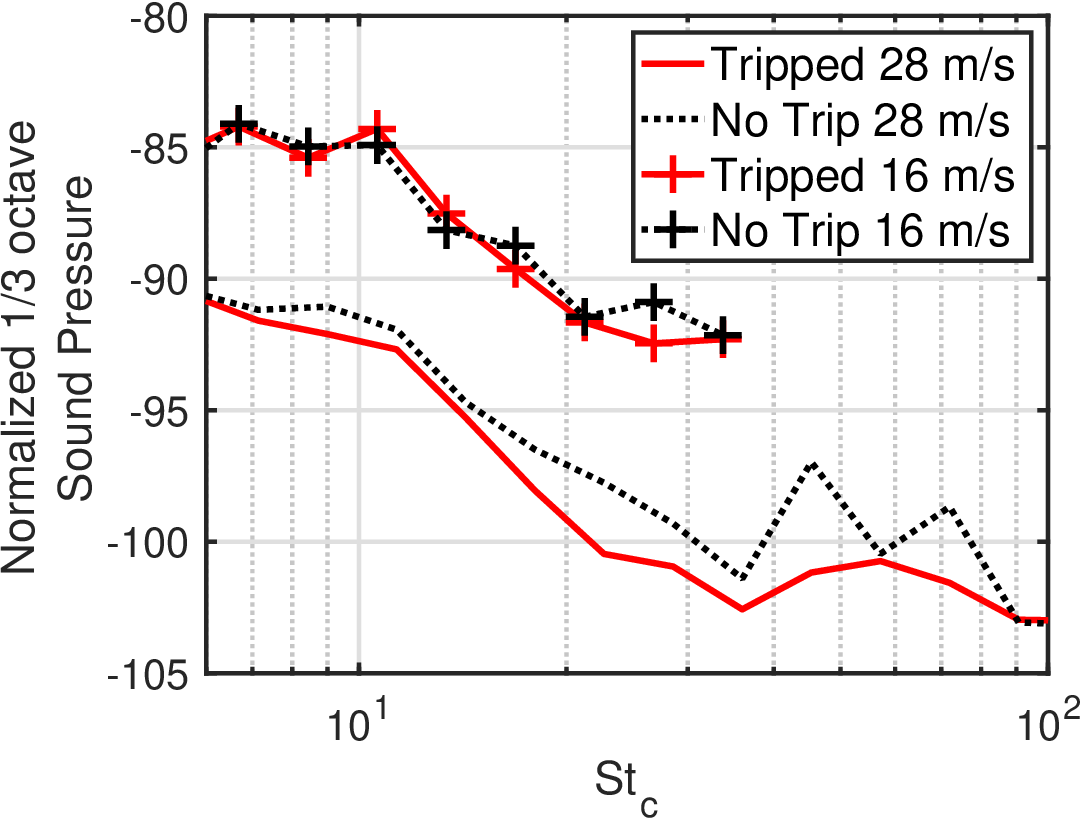} \\
  \end{tabular}

\caption{Normalized third-octave band sound pressure as function of $St_C$ (a) Non compact dipole scaling, $U_{\infty}^{4}$; (b) jet noise scaling, $U_{\infty}^{7}$.} 
\label{Sppnorm} 
\end{figure*}

As already shown in figure \ref{Acoustic_spectra}, tip noise operates at high frequencies (1000 Hz) where sources are non-compact relative to airfoil chord, Helmholtz numbers $kC > 2$, where $k$ is the acoustic wavenumber. As such, the applicability of the scaling proposed by \citet{williams1970aerodynamic} for non-compact sources is tested by plotting the acoustic pressure of the third octave band normalized by the fourth power of the free-stream velocity. The normalized third-octave band pressure ($10\times\log_{10}\bigg[\frac{S_{pp} \times R^2}{U^4 \times t_p^2 C}\bigg]$) is plotted in figure \ref{Sppnorm} (\textit{a}) as a function of the Strouhal number based on chord length and free-stream velocity (St$_C$). The nominal distance, $\frac{R^2}{t_p^2 C}$, was used for asymptotic scaling laws. Note that normalization with $U^4$ in St$_C$ corresponds to $U^5$ scaling in frequency. A good collapse is obtained for St$_C$ less than 20. At higher frequencies, $U_{\infty}^5$ scaling fails. More importantly, the proposed $U_{\infty}^8$ quadrupole scaling \cite{Grilliat2007,jacob2010aeroacoustic,moreau2016experimental} in frequency, and thus $U_{\infty}^7$ in $St_c$, shown in figure \ref{Sppnorm} (\textit{b}), also fails to achieve collapse at high frequencies.

\begin{figure*}
  \centering
  \begin{tabular}{@{}p{0.45\linewidth}@{\quad}p{0.45\linewidth}@{}}
    \subfigimg[width=70 mm,pos=ul,vsep=21pt,hsep=42pt]{(a)}{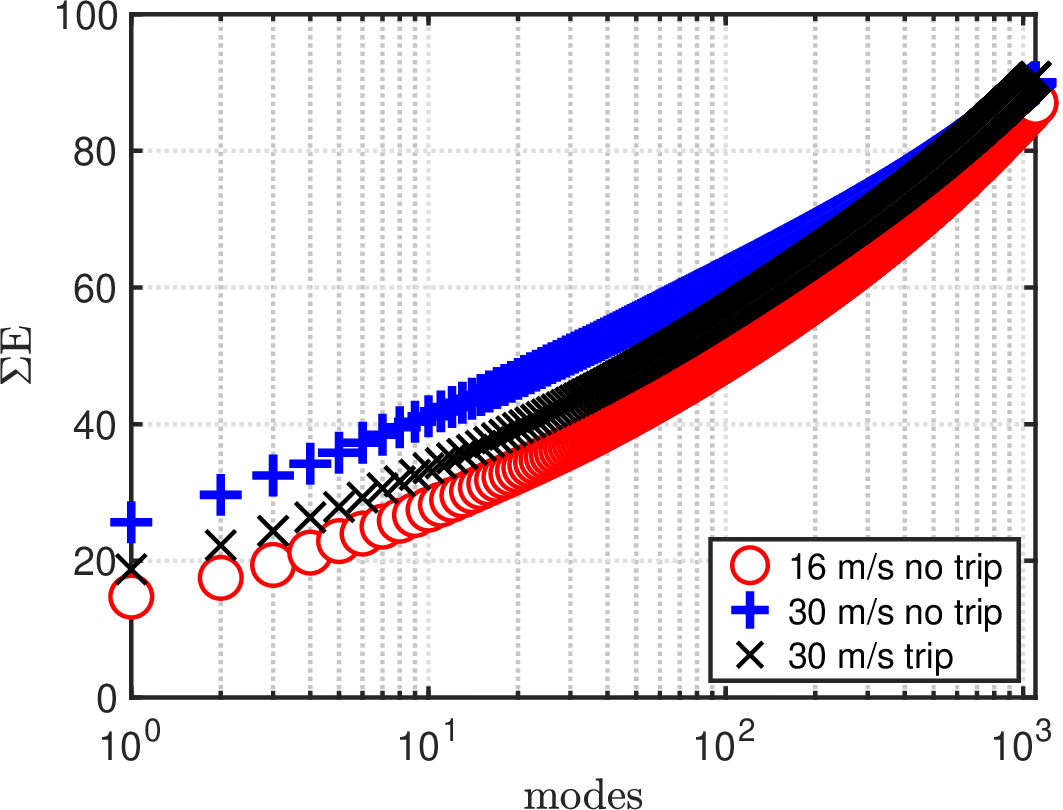} &
    \subfigimg[width=70 mm,pos=ur,vsep=15pt,hsep=22pt]{(b)}{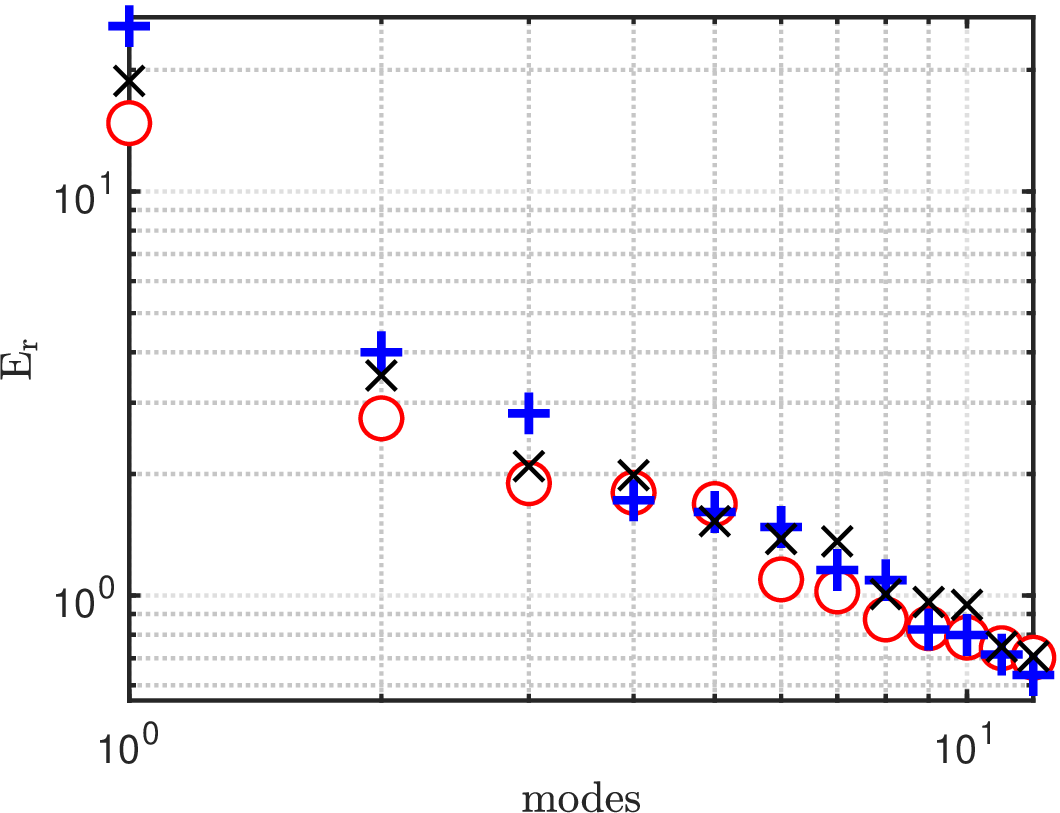} \\
  \end{tabular}

\caption{POD of the measured velocity field in the trailing-edge region (camera 2). (a) Cumulative sum of POD modes; (b) Relative energy of individual POD modes.} 
\label{Energy_modes} 
\end{figure*}

To unravel the modes in the velocity disturbance field, a Proper Orthogonal Decomposition (POD) 
was achieved 
using the snapshot approach 
developed by \citet{sirovich1987turbulence}. Since, turbulence intensity in the leading-edge region is close to zero, the POD analysis was performed only in the trailing-edge region of the airfoil. The resulting distribution of energy, for example the eigenvalues of the POD analysis, is shown in Figure \ref{Energy_modes}. A large part of the total kinetic energy is present within a small number of modes, typical of flows with convective coherent structures. For the rough case, modes 7 and 8 constitute a modal pair, while for the smooth-wall case, modes 8 and 9 comprise a modal pair. Since, the differences between the two smooth cases, at 16 and 28 m/s, are minor no further comparative analysis will be performed between the two cases.

\begin{figure*}
  \centering
  \begin{tabular}{@{}p{0.45\linewidth}@{\quad}p{0.45\linewidth}@{}}
    \subfigimg[width=70 mm,pos=ll,vsep=21pt,hsep=26pt]{(a)}{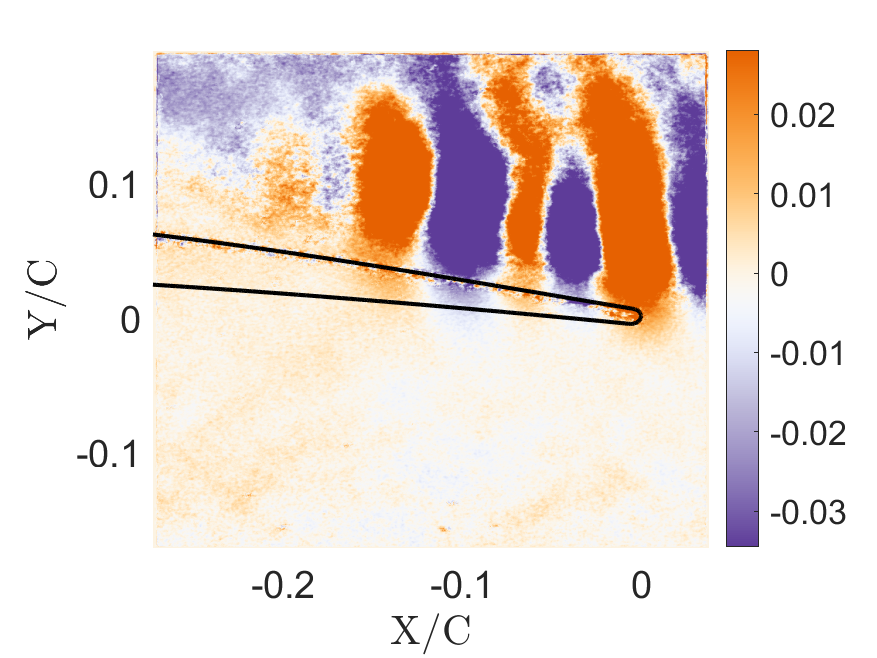} &
    \subfigimg[width=70 mm,pos=ll,vsep=21pt,hsep=26pt]{(b)}{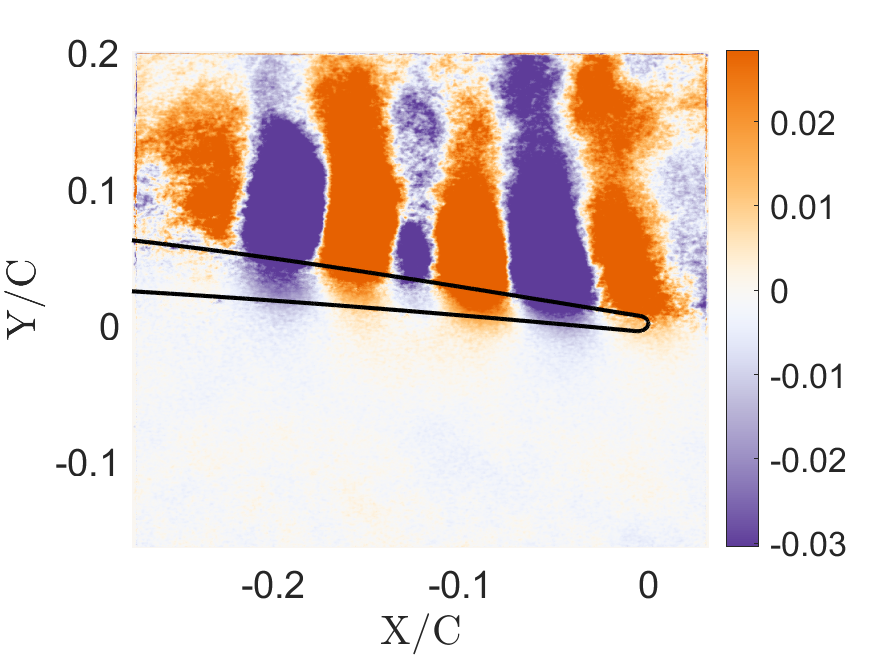} \\
  \end{tabular}

\caption{POD mode 8 of the vertical velocity disturbance field at $U_{\infty}=28$ m/s; (a) smooth, (b) rough cases.} 
\label{spatial_modes} 
\end{figure*}

Figure \ref{spatial_modes} shows the spatial mode 8 of the vertical velocity disturbance field for $U_{\infty}=28$ m/s. This particular mode has been chosen for further analysis because it features as a convective mode in either of the rough or smooth wall measurements, and it has comparable energy content in these two cases. The spatial modes correspond to the spatial organization of the velocity disturbance field in space, i.e. coherent structures. The spatial organization of this particular mode is distributed across the trailing-edge for the rough case, while in the smooth case the modal disturbance of coherent structure is more localized away from the airfoil, which is consistent with our observations on the standard deviation of velocity. These modes are predominantly located on the airfoil suction side and have been shown to be associated with tip leakage noise and flow instability by \citet{saraceno2022tip} and \citet{Koch_diss}. 

\begin{figure*}
  \centering
  \begin{tabular}{@{}p{0.45\linewidth}@{\quad}p{0.45\linewidth}@{}}
    \subfigimg[width=70 mm,pos=ur,vsep=12pt,hsep=12pt]{(a)}{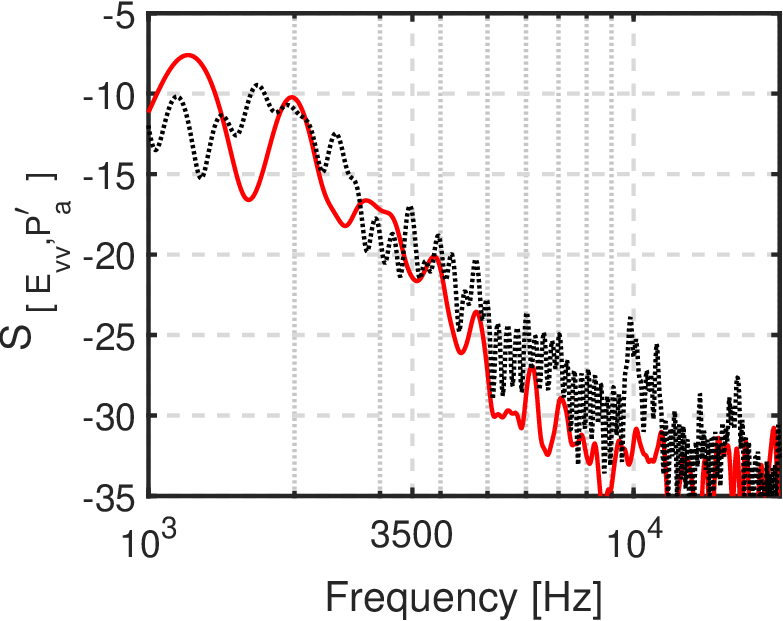} &
    \subfigimg[width=70 mm,pos=ur,vsep=15pt,hsep=22pt]{(b)}{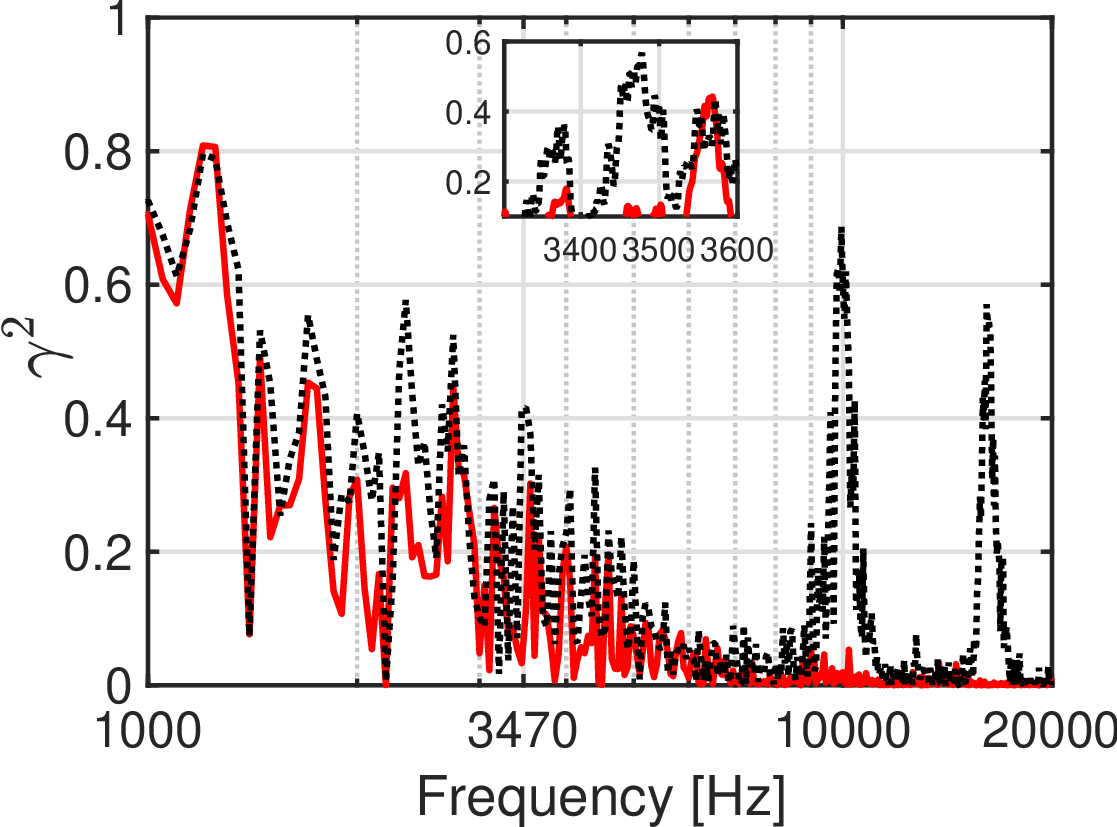} \\
  \end{tabular}

\caption{(a) Magnitude of the cross-spectral density between mode 8 and acoustic pressure; (b) coherence between far-field microphones. Legends: the same as in Figure \ref{Acoustic_spectra}.} 
\label{spectra_modes} 
\end{figure*}

Previous studies \citep{camussi2010experimental,saraceno2022tip} have tried to establish a causal link between these flow instabilities and acoustics through correlation; ultimately, causality can be more reasonably linked through interventions~\citep{pearl2000models}. Therefore, the correlations between acoustics and near-field flow disturbances, with and without roughness, are compared because roughness acts as an intervention in far-field tonal noise. 
The cross-correlation between mode 8 and far-field acoustics was computed, the spectral content of which was estimated using the Welch's method. The resulting quantity is shown in figure \ref{spectra_modes} (\textit{a}) for the two cases. Distinct peaks at about 2500, 3500, 10000 and 16000~~Hz can be seen for the case with no roughness, which confirms that mode 8 of the smooth-wall case does indeed carry information on the acoustic spectral peaks observed 
in the noise measurements. In fact, the additional peak ($\approx 2500$~Hz) emerges from the magnitude-squared coherence between two microphones placed on either side of the airfoil in figure \ref{spectra_modes} (\textit{b}). Moreover, strong coherence is also observed for the two newly observed humps at high frequencies. This further strengthens the observations made in figure \ref{spectra_modes} (\textit{a}). As in the SPL in figure~\ref{Acoustic_spectra}, all theses peaks disappear with roughness.

The coherence between the pressure and suction sides of the airfoil can also be elucidated using the two-point zero time delay correlation of the cross-flow velocity disturbance $V'$, defined as:

{\begin{equation} 
\textcolor{black}{R_{V'V'}({x},{x}^{\prime},y,{y}^{\prime},z,{z}^{\prime}) =   \frac{\overline{{V'}(x,y,z) {V'}({x}^{\prime},{y}^{\prime},{z}^{\prime}) }}{\sqrt{{\overline{V'(x,y,z})^2}} \times \sqrt{\overline{{V'}({x}^{\prime},{y}^{\prime},{z}^{\prime})^2}}}}
  \label{two_point}         
\end{equation}}

To this end, the fixed point was selected based on the location where the maximum covariance between the far-field pressure and vertical velocity disturbances was observed. Subsequently, the two-point correlation was computed, as shown in figure \ref{two_point} for the two cases. The location of the fixed-point has been highlighted as a cross in the two images. Overall, the correlation of the cross-flow velocity disturbances remains elevated in the tip-gap region. However, the addition of roughness, figure \ref{two_point} (\textit{b}), clearly weakens the correlation over the suction side of the airfoil, especially near the trailing edge region $X/C \ge -0.2$, thus diminishing the coherence on either side of the airfoil tip.

\begin{figure*}
  \centering
  \begin{tabular}{@{}p{0.45\linewidth}@{\quad}p{0.45\linewidth}@{}}
    \subfigimg[width=70 mm,pos=ll,vsep=30pt,hsep=38pt]{(a)}{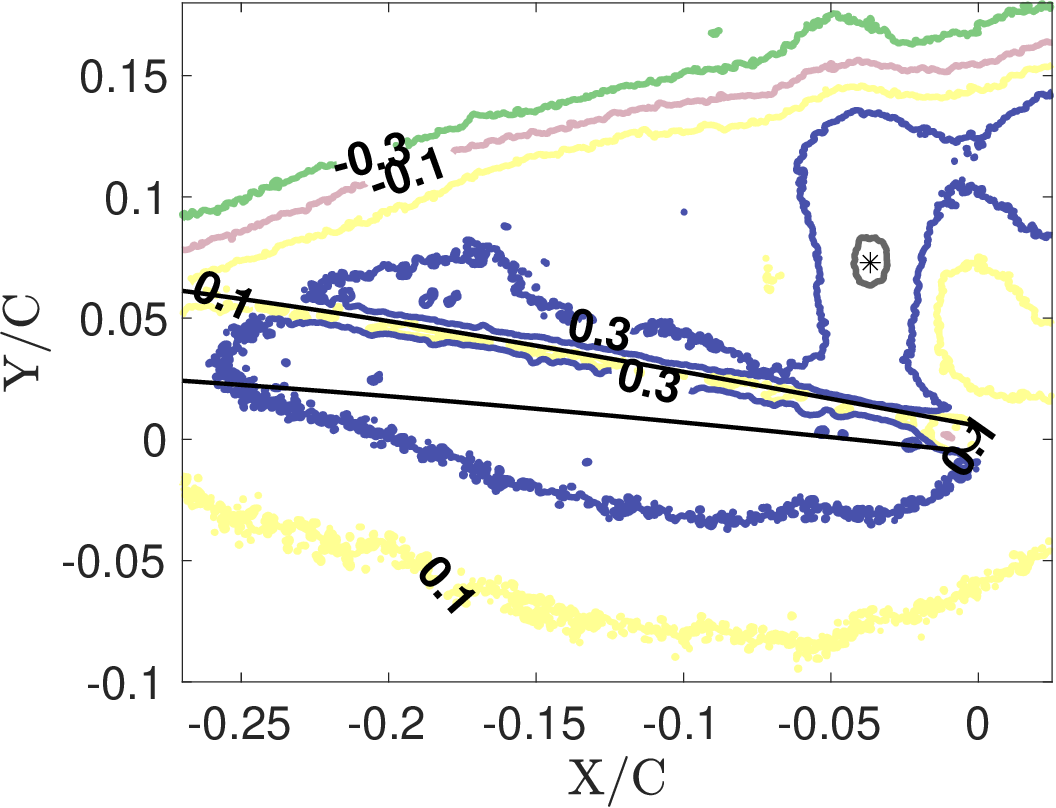} &
    \subfigimg[width=70 mm,pos=ll,vsep=30pt,hsep=38pt]{(b)}{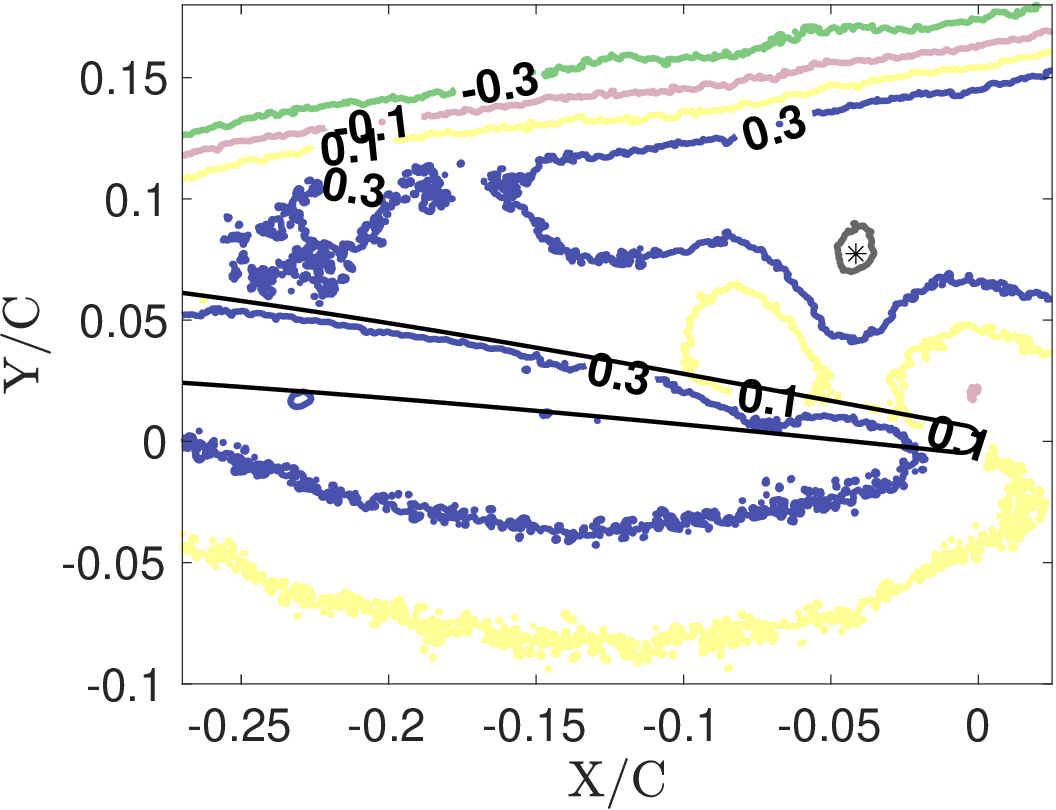} \\
  \end{tabular}

\caption{Two-point zero time-delay correlation of the vertical velocity disturbance field at $U_{\infty}=28$ m/s; (a) smooth, (b) rough cases. The crosses in the two plots correspond to the location of the fixed point, X/C$\approx -0.04$ and Y/C $\approx 0.07$.} 
\label{two_point} 
\end{figure*}

\section{Discussion} \label{sec:disc}

Introducing surface roughness on the pressure side of the airfoil alters cross-flow velocity, with increased cross-stream velocity near the mid-chord and decreased velocity close to the trailing edge. 
  These differences most likely result from secondary flows over the pressure side due to surface roughness \citep[see][for instance]{chung2018similarity}. This allows us to compare differences in far-field acoustics due to weaker cross flow ($V/U_{\infty}$) at matched Reynolds and Mach numbers. Together with the lower background noise (jet noise), the absence of flow separation in the trailing edge region, and manipulation of the cross-flow velocity help to elucidate some of the key noise mechanisms behind tip leakage noise from a stationary airfoil.

The present study reveals for the first time that the tones associated with the tip gap extend beyond previously reported frequencies, that is, 3.5 kHz \citep{jacob2016time,saraceno2022tip}. High frequency peaks have not been reported in previous measurements, perhaps because the current set of measurements is performed at lower Mach numbers compared to previous studies \citep{camussi2010experimental,jacob2016time,saraceno2022tip,palleja2022reduction}, leading to a much lower contribution of background jet noise in the present study. Furthermore, previous studies \citep{camussi2010experimental,jacob2016time,saraceno2022tip,palleja2022reduction} are performed at a very high angle of attack, where flow separation is expected on the suction side of the airfoil.


Modal shapes confirm that instabilities in the tip-gap region for the rough-wall case start upstream compared to the smooth case, consistent with a more localized region of elevated cross flow velocity for the smooth-wall case. Coherent structures are more confined for the smooth-wall case, carrying information associated with far-field noise peaks absent in the rough-wall case. Importantly, while the negative correlation between far-field microphones at low and mid frequencies indicates a dipolar character of the noise sources, at higher frequencies (e.g. beyond 10 kHz), the noise sources exhibit a non-dipolar nature and are consistent with the inability of $U_{\infty}^5$ to scale acoustic pressure at these high frequencies. However, far-field acoustics also do not follow the $U_{\infty}^8$ scaling, which implies that the high-frequency sound radiation mechanism, St$_c \ge 20$, cannot be uniquely attributed to jet-like noise sources and is thus in contradiction to the findings in \cite{Grilliat2007,jacob2010aeroacoustic}. This can be attributed to the higher background noise (jet noise), in previous experiments \cite{Grilliat2007,jacob2010aeroacoustic}, which can affect the high-frequency noise.

Although modal shapes and the cross-spectral density between modes and far-field acoustics can imply that coherent structures associated with flow instabilities can cause tip leakage noise \citep{camussi2010experimental,saraceno2022tip}, 
  correlation alone does not imply causation, and this is demonstrated by the fact that these coherent structures are present in either case. If 
  the intervention route is taken and 
  the effects of roughness on noise are compared, 
  roughness on the pressure side is seen to lead to cancellation of spectral peaks and a drop in the cross-spectral density at the frequencies of tones. Roughness leads to a reduction in the velocity and acoustic pressure correlation pattern between airfoil pressure and suction side. The origin of acoustic peaks is then linked to coherence between the airfoil sides near the tip. For airfoil-tip with surface roughness, a reduction in high-frequency noise achieved despite elevated region of velocity fluctuations induced by the surface-roughness in the tip-gap region further gives credence to the high-frequency tones being linked to coherence between the airfoil suction and pressure sides. 
  
  As porous walls can be considered rough \citep{jaiswal2023effects}, the absence of coherence between the two sides of the airfoil through the introduction of roughness elements or roughness induced by porous surfaces can suppress these peaks in tip leakage noise. For the porous wall boundary condition, \citet{palleja2022reduction} have reported the absence of coherence between the wall pressure and the far field. The added requirement of coherence between the pressure and suction sides of the airfoil implies that an acoustic feedback loop 
  or a cross-communication 
  could potentially explain this frequency selection process, which is removed by roughness introduced on the pressure-side edge. This is also consistent with the findings of \citet{Moreau2016}, who also showed a reduction in airfoil tonal noise using porous liner-like materials. It is important to note that the causal link between noise peaks and wall boundary conditions is established here not by correlations, but rather through intervention~\citep{pearl2000models}. Here, the reader is cautioned that the existence of a smooth wall boundary condition on either side of the airfoil tip may still not be a sufficient condition for the existence of these tones.~In the present set of experiments, only the temperature (hence the speed of sound), the airfoil geometry and the tip gap were kept constant. Therefore, it is possible that the thickness of the airfoil and the speed of the sound play an important role in the frequency selection process of these peaks. However, future studies in this direction are needed to further understand these withstanding issues.  

  \section{Conclusion} \label{sec:conclusion}

    Velocity and noise measurements for tip-leakage flows over a fixed airfoil have been conducted. Two research questions have guided this study: 1) What is the operational range of tip-gap noise? 2) What is the origin of humps or tones in the acoustic spectra, and can they be solely attributed to coherent structures in the tip-gap region?
  
  In a low-noise facility, 
  tip-leakage noise is found to have two peaks and humps at high frequencies, around 10 and 16.5~kHz, which have been measured for the first time. Several experiments, including our own, have shown the existence of high-amplitude acoustic peaks emitted by a stationary airfoil with a tip gap. More importantly, these acoustic tones do not follow any classical Strouhal or Reynolds number scalings.

  Given differences in Reynolds and Mach numbers and key geometry differences (airfoil, side plate height, and size of the tip gap), these peaks cannot therefore be linked to acoustics or aerodynamics alone, because the existence of flow instabilities alone does not ensure frequency selection of these tones; instead, coherence in velocity and acoustic perturbations on either side of the airfoil tip is required, which itself requires a smooth wall boundary condition on either side of the airfoil. Therefore, the introduction of surface roughness at the airfoil tip on the pressure side reduces high-frequency noise and mitigates acoustic peaks, indicating the feasibility of surface roughness as a noise mitigation strategy. Consequently, surface roughness holds promise in reducing tip leakage noise in ducted fans and propellers, which warrants further investigation.
  
  \begin{acknowledgments}
PJ acknowledges the financial support from UK Research and Innovation (UKRI) under the UK government’s Horizon Europe funding guarantee [grant number EP/X032590/1)]. 
\end{acknowledgments}


%






\bibliography{apssamp}

\begin{thebibliography}{48}%
\makeatletter
\providecommand \@ifxundefined [1]{%
 \@ifx{#1\undefined}
}%
\providecommand \@ifnum [1]{%
 \ifnum #1\expandafter \@firstoftwo
 \else \expandafter \@secondoftwo
 \fi
}%
\providecommand \@ifx [1]{%
 \ifx #1\expandafter \@firstoftwo
 \else \expandafter \@secondoftwo
 \fi
}%
\providecommand \natexlab [1]{#1}%
\providecommand \enquote  [1]{``#1''}%
\providecommand \bibnamefont  [1]{#1}%
\providecommand \bibfnamefont [1]{#1}%
\providecommand \citenamefont [1]{#1}%
\providecommand \href@noop [0]{\@secondoftwo}%
\providecommand \href [0]{\begingroup \@sanitize@url \@href}%
\providecommand \@href[1]{\@@startlink{#1}\@@href}%
\providecommand \@@href[1]{\endgroup#1\@@endlink}%
\providecommand \@sanitize@url [0]{\catcode `\\12\catcode `\$12\catcode `\&12\catcode `\#12\catcode `\^12\catcode `\_12\catcode `\%12\relax}%
\providecommand \@@startlink[1]{}%
\providecommand \@@endlink[0]{}%
\providecommand \url  [0]{\begingroup\@sanitize@url \@url }%
\providecommand \@url [1]{\endgroup\@href {#1}{\urlprefix }}%
\providecommand \urlprefix  [0]{URL }%
\providecommand \Eprint [0]{\href }%
\providecommand \doibase [0]{https://doi.org/}%
\providecommand \selectlanguage [0]{\@gobble}%
\providecommand \bibinfo  [0]{\@secondoftwo}%
\providecommand \bibfield  [0]{\@secondoftwo}%
\providecommand \translation [1]{[#1]}%
\providecommand \BibitemOpen [0]{}%
\providecommand \bibitemStop [0]{}%
\providecommand \bibitemNoStop [0]{.\EOS\space}%
\providecommand \EOS [0]{\spacefactor3000\relax}%
\providecommand \BibitemShut  [1]{\csname bibitem#1\endcsname}%
\let\auto@bib@innerbib\@empty
\bibitem [{\citenamefont {Moreau}\ and\ \citenamefont {Roger}(2018)}]{moreau2018advanced}%
  \BibitemOpen
  \bibfield  {author} {\bibinfo {author} {\bibfnamefont {S.}~\bibnamefont {Moreau}}\ and\ \bibinfo {author} {\bibfnamefont {M.}~\bibnamefont {Roger}},\ }\bibfield  {title} {\bibinfo {title} {Advanced noise modeling for future propulsion systems},\ }\href@noop {} {\bibfield  {journal} {\bibinfo  {journal} {Int. J. of Aeroacoustics}\ }\textbf {\bibinfo {volume} {17}},\ \bibinfo {pages} {576} (\bibinfo {year} {2018})}\BibitemShut {NoStop}%
\bibitem [{\citenamefont {Moreau}(2019)}]{moreau2019turbomachinery}%
  \BibitemOpen
  \bibfield  {author} {\bibinfo {author} {\bibfnamefont {S.}~\bibnamefont {Moreau}},\ }\bibfield  {title} {\bibinfo {title} {Turbomachinery noise predictions: present and future},\ }\href@noop {} {\bibfield  {journal} {\bibinfo  {journal} {Acoustics}\ }\textbf {\bibinfo {volume} {1}},\ \bibinfo {pages} {92} (\bibinfo {year} {2019})}\BibitemShut {NoStop}%
\bibitem [{\citenamefont {Moreau}\ and\ \citenamefont {Roger}(2024)}]{Moreau2024}%
  \BibitemOpen
  \bibfield  {author} {\bibinfo {author} {\bibfnamefont {S.}~\bibnamefont {Moreau}}\ and\ \bibinfo {author} {\bibfnamefont {M.}~\bibnamefont {Roger}},\ }\bibfield  {title} {\bibinfo {title} {{Turbomachinery noise review}},\ }\href@noop {} {\bibfield  {journal} {\bibinfo  {journal} {Int. J. Turbomach. Propuls. Power}\ }\textbf {\bibinfo {volume} {9}},\ \bibinfo {pages} {1} (\bibinfo {year} {2024})}\BibitemShut {NoStop}%
\bibitem [{\citenamefont {Moreau}\ and\ \citenamefont {Sanjos{\'e}}(2016)}]{Moreau:2016}%
  \BibitemOpen
  \bibfield  {author} {\bibinfo {author} {\bibfnamefont {S.}~\bibnamefont {Moreau}}\ and\ \bibinfo {author} {\bibfnamefont {M.}~\bibnamefont {Sanjos{\'e}}},\ }\bibfield  {title} {\bibinfo {title} {{Sub-harmonic broadband humps and tip noise in low-speed ring fans}},\ }\href {https://doi.org/10.1121/1.4939493} {\bibfield  {journal} {\bibinfo  {journal} {{J. Acoust. Soc. Am.}}\ }\textbf {\bibinfo {volume} {139}},\ \bibinfo {pages} {118} (\bibinfo {year} {2016})}\BibitemShut {NoStop}%
\bibitem [{\citenamefont {Grilliat}\ \emph {et~al.}(2007)\citenamefont {Grilliat}, \citenamefont {Jacob}, \citenamefont {Camussi},\ and\ \citenamefont {Caputi-Gennaro}}]{Grilliat2007}%
  \BibitemOpen
  \bibfield  {author} {\bibinfo {author} {\bibfnamefont {J.}~\bibnamefont {Grilliat}}, \bibinfo {author} {\bibfnamefont {M.~C.}\ \bibnamefont {Jacob}}, \bibinfo {author} {\bibfnamefont {R.}~\bibnamefont {Camussi}},\ and\ \bibinfo {author} {\bibfnamefont {G.}~\bibnamefont {Caputi-Gennaro}},\ }\bibfield  {title} {\bibinfo {title} {{Tip leakage experiment - Part one: Aerodynamic and acoustic measurements}},\ }\href {https://doi.org/10.2514/6.2007-3684} {\bibfield  {journal} {\bibinfo  {journal} {13th AIAA/CEAS Aeroacoustics Conference (28th AIAA Aeroacoustics Conference), paper}\ ,\ \bibinfo {pages} {1}} (\bibinfo {year} {2007})}\BibitemShut {NoStop}%
\bibitem [{\citenamefont {Camussi}\ \emph {et~al.}(2010)\citenamefont {Camussi}, \citenamefont {Grilliat}, \citenamefont {Caputi-Gennaro},\ and\ \citenamefont {Jacob}}]{camussi2010experimental}%
  \BibitemOpen
  \bibfield  {author} {\bibinfo {author} {\bibfnamefont {R.}~\bibnamefont {Camussi}}, \bibinfo {author} {\bibfnamefont {J.}~\bibnamefont {Grilliat}}, \bibinfo {author} {\bibfnamefont {G.}~\bibnamefont {Caputi-Gennaro}},\ and\ \bibinfo {author} {\bibfnamefont {M.~C.}\ \bibnamefont {Jacob}},\ }\bibfield  {title} {\bibinfo {title} {Experimental study of a tip leakage flow: wavelet analysis of pressure fluctuations},\ }\href@noop {} {\bibfield  {journal} {\bibinfo  {journal} {J. Fluid Mech.}\ }\textbf {\bibinfo {volume} {660}},\ \bibinfo {pages} {87} (\bibinfo {year} {2010})}\BibitemShut {NoStop}%
\bibitem [{\citenamefont {Jacob}\ \emph {et~al.}(2016)\citenamefont {Jacob}, \citenamefont {Jondeau},\ and\ \citenamefont {Li}}]{jacob2016time}%
  \BibitemOpen
  \bibfield  {author} {\bibinfo {author} {\bibfnamefont {M.~C.}\ \bibnamefont {Jacob}}, \bibinfo {author} {\bibfnamefont {E.}~\bibnamefont {Jondeau}},\ and\ \bibinfo {author} {\bibfnamefont {B.}~\bibnamefont {Li}},\ }\bibfield  {title} {\bibinfo {title} {Time-resolved piv measurements of a tip leakage flow},\ }\href@noop {} {\bibfield  {journal} {\bibinfo  {journal} {Int. J. of Aeroacoustics}\ }\textbf {\bibinfo {volume} {15}},\ \bibinfo {pages} {662} (\bibinfo {year} {2016})}\BibitemShut {NoStop}%
\bibitem [{\citenamefont {Saraceno}\ \emph {et~al.}(2022)\citenamefont {Saraceno}, \citenamefont {Palleja-Cabre}, \citenamefont {Jaiswal}, \citenamefont {Paruchuri},\ and\ \citenamefont {Ganapathisubramani}}]{saraceno2022tip}%
  \BibitemOpen
  \bibfield  {author} {\bibinfo {author} {\bibfnamefont {I.}~\bibnamefont {Saraceno}}, \bibinfo {author} {\bibfnamefont {S.}~\bibnamefont {Palleja-Cabre}}, \bibinfo {author} {\bibfnamefont {P.}~\bibnamefont {Jaiswal}}, \bibinfo {author} {\bibfnamefont {C.~C.}\ \bibnamefont {Paruchuri}},\ and\ \bibinfo {author} {\bibfnamefont {B.}~\bibnamefont {Ganapathisubramani}},\ }\bibfield  {title} {\bibinfo {title} {On the tip leakage noise generating mechanisms of single-fixed aerofoil},\ }\href@noop {} {\bibfield  {journal} {\bibinfo  {journal} {28th AIAA/CEAS Aeroacoustics 2022 Conference, Paper number 2881}\ } (\bibinfo {year} {2022})}\BibitemShut {NoStop}%
\bibitem [{\citenamefont {Furukawa}\ \emph {et~al.}(1998)\citenamefont {Furukawa}, \citenamefont {Saiki}, \citenamefont {Nagayoshi}, \citenamefont {Kuroumaru},\ and\ \citenamefont {Inoue}}]{furukawa1998effects}%
  \BibitemOpen
  \bibfield  {author} {\bibinfo {author} {\bibfnamefont {M.}~\bibnamefont {Furukawa}}, \bibinfo {author} {\bibfnamefont {K.}~\bibnamefont {Saiki}}, \bibinfo {author} {\bibfnamefont {K.}~\bibnamefont {Nagayoshi}}, \bibinfo {author} {\bibfnamefont {M.}~\bibnamefont {Kuroumaru}},\ and\ \bibinfo {author} {\bibfnamefont {M.}~\bibnamefont {Inoue}},\ }\bibfield  {title} {\bibinfo {title} {Effects of stream surface inclination on tip leakage flow fields in compressor rotors},\ }\href@noop {} {\bibfield  {journal} {\bibinfo  {journal} {J. of Turbomachinery}\ }\textbf {\bibinfo {volume} {120}},\ \bibinfo {pages} {683} (\bibinfo {year} {1998})}\BibitemShut {NoStop}%
\bibitem [{\citenamefont {Goto}(1992)}]{goto1992three}%
  \BibitemOpen
  \bibfield  {author} {\bibinfo {author} {\bibfnamefont {A.}~\bibnamefont {Goto}},\ }\bibfield  {title} {\bibinfo {title} {Three-dimensional flow and mixing in an axial flow compressor with different rotor tip clearances},\ }\href@noop {} {\bibfield  {journal} {\bibinfo  {journal} {J. of Turbomachinery}\ }\textbf {\bibinfo {volume} {114}},\ \bibinfo {pages} {675} (\bibinfo {year} {1992})}\BibitemShut {NoStop}%
\bibitem [{\citenamefont {Muthanna}\ and\ \citenamefont {Devenport}(2004)}]{muthanna2004wake}%
  \BibitemOpen
  \bibfield  {author} {\bibinfo {author} {\bibfnamefont {C.}~\bibnamefont {Muthanna}}\ and\ \bibinfo {author} {\bibfnamefont {W.~J.}\ \bibnamefont {Devenport}},\ }\bibfield  {title} {\bibinfo {title} {Wake of a compressor cascade with tip gap, part 1: Mean flow and turbulence structure},\ }\href@noop {} {\bibfield  {journal} {\bibinfo  {journal} {AIAA J.}\ }\textbf {\bibinfo {volume} {42}},\ \bibinfo {pages} {2320} (\bibinfo {year} {2004})}\BibitemShut {NoStop}%
\bibitem [{\citenamefont {Koch}\ \emph {et~al.}(2021)\citenamefont {Koch}, \citenamefont {Sanjos{\'e}},\ and\ \citenamefont {Moreau}}]{koch2021large}%
  \BibitemOpen
  \bibfield  {author} {\bibinfo {author} {\bibfnamefont {R.}~\bibnamefont {Koch}}, \bibinfo {author} {\bibfnamefont {M.}~\bibnamefont {Sanjos{\'e}}},\ and\ \bibinfo {author} {\bibfnamefont {S.}~\bibnamefont {Moreau}},\ }\bibfield  {title} {\bibinfo {title} {Large-eddy simulation of a single airfoil tip-leakage flow},\ }\href@noop {} {\bibfield  {journal} {\bibinfo  {journal} {AIAA J.}\ }\textbf {\bibinfo {volume} {59}},\ \bibinfo {pages} {2546} (\bibinfo {year} {2021})}\BibitemShut {NoStop}%
\bibitem [{\citenamefont {Jacob}\ \emph {et~al.}(2010)\citenamefont {Jacob}, \citenamefont {Grilliat}, \citenamefont {Camussi},\ and\ \citenamefont {Gennaro}}]{jacob2010aeroacoustic}%
  \BibitemOpen
  \bibfield  {author} {\bibinfo {author} {\bibfnamefont {M.~C.}\ \bibnamefont {Jacob}}, \bibinfo {author} {\bibfnamefont {J.}~\bibnamefont {Grilliat}}, \bibinfo {author} {\bibfnamefont {R.}~\bibnamefont {Camussi}},\ and\ \bibinfo {author} {\bibfnamefont {G.~C.}\ \bibnamefont {Gennaro}},\ }\bibfield  {title} {\bibinfo {title} {Aeroacoustic investigation of a single airfoil tip leakage flow},\ }\href@noop {} {\bibfield  {journal} {\bibinfo  {journal} {International Journal of Aeroacoustics}\ }\textbf {\bibinfo {volume} {9}},\ \bibinfo {pages} {253} (\bibinfo {year} {2010})}\BibitemShut {NoStop}%
\bibitem [{\citenamefont {Lighthill}(1952)}]{lighthill1952sound}%
  \BibitemOpen
  \bibfield  {author} {\bibinfo {author} {\bibfnamefont {M.~J.}\ \bibnamefont {Lighthill}},\ }\bibfield  {title} {\bibinfo {title} {On sound generated aerodynamically i. general theory},\ }\href@noop {} {\bibfield  {journal} {\bibinfo  {journal} {Proceedings of the Royal Society of London. Series A. Mathematical and Physical Sciences}\ }\textbf {\bibinfo {volume} {211}},\ \bibinfo {pages} {564} (\bibinfo {year} {1952})}\BibitemShut {NoStop}%
\bibitem [{\citenamefont {Boudet}\ \emph {et~al.}(2016)\citenamefont {Boudet}, \citenamefont {Caro}, \citenamefont {Li}, \citenamefont {Jondeau},\ and\ \citenamefont {Jacob}}]{boudet2016zonal}%
  \BibitemOpen
  \bibfield  {author} {\bibinfo {author} {\bibfnamefont {J.}~\bibnamefont {Boudet}}, \bibinfo {author} {\bibfnamefont {J.}~\bibnamefont {Caro}}, \bibinfo {author} {\bibfnamefont {B.}~\bibnamefont {Li}}, \bibinfo {author} {\bibfnamefont {E.}~\bibnamefont {Jondeau}},\ and\ \bibinfo {author} {\bibfnamefont {M.~C.}\ \bibnamefont {Jacob}},\ }\bibfield  {title} {\bibinfo {title} {Zonal large-eddy simulation of a tip leakage flow},\ }\href@noop {} {\bibfield  {journal} {\bibinfo  {journal} {Int. J. of Aeroacoustics}\ }\textbf {\bibinfo {volume} {15}},\ \bibinfo {pages} {646} (\bibinfo {year} {2016})}\BibitemShut {NoStop}%
\bibitem [{\citenamefont {Moreau}\ \emph {et~al.}(2016{\natexlab{a}})\citenamefont {Moreau}, \citenamefont {Doolan}, \citenamefont {Alexander}, \citenamefont {Meyers},\ and\ \citenamefont {Devenport}}]{moreau2016wall}%
  \BibitemOpen
  \bibfield  {author} {\bibinfo {author} {\bibfnamefont {D.~J.}\ \bibnamefont {Moreau}}, \bibinfo {author} {\bibfnamefont {C.~J.}\ \bibnamefont {Doolan}}, \bibinfo {author} {\bibfnamefont {W.~N.}\ \bibnamefont {Alexander}}, \bibinfo {author} {\bibfnamefont {T.~W.}\ \bibnamefont {Meyers}},\ and\ \bibinfo {author} {\bibfnamefont {W.~J.}\ \bibnamefont {Devenport}},\ }\bibfield  {title} {\bibinfo {title} {Wall-mounted finite airfoil-noise production and prediction},\ }\href@noop {} {\bibfield  {journal} {\bibinfo  {journal} {AIAA Journal}\ }\textbf {\bibinfo {volume} {54}},\ \bibinfo {pages} {1637} (\bibinfo {year} {2016}{\natexlab{a}})}\BibitemShut {NoStop}%
\bibitem [{\citenamefont {Rossiter}(1964)}]{rossiter1964wind}%
  \BibitemOpen
  \bibfield  {author} {\bibinfo {author} {\bibfnamefont {J.}~\bibnamefont {Rossiter}},\ }\bibfield  {title} {\bibinfo {title} {Wind-tunnel experiments on the flow over rectangular cavities at subsonic and transonic speeds},\ }\href@noop {} {\bibfield  {journal} {\bibinfo  {journal} {RAE Technical Report}\ } (\bibinfo {year} {1964})}\BibitemShut {NoStop}%
\bibitem [{\citenamefont {Sanjos\'e}\ \emph {et~al.}(2019)\citenamefont {Sanjos\'e}, \citenamefont {Towne}, \citenamefont {Jaiswal}, \citenamefont {Moreau},\ and\ \citenamefont {Lele}}]{Sanjose2019}%
  \BibitemOpen
  \bibfield  {author} {\bibinfo {author} {\bibfnamefont {M.}~\bibnamefont {Sanjos\'e}}, \bibinfo {author} {\bibfnamefont {A.}~\bibnamefont {Towne}}, \bibinfo {author} {\bibfnamefont {P.}~\bibnamefont {Jaiswal}}, \bibinfo {author} {\bibfnamefont {S.}~\bibnamefont {Moreau}},\ and\ \bibinfo {author} {\bibfnamefont {S.}~\bibnamefont {Lele}},\ }\bibfield  {title} {\bibinfo {title} {{Modal analysis of the laminar boundary layer instability and tonal noise of an airfoil at Reynolds number 150,000}},\ }\href@noop {} {\bibfield  {journal} {\bibinfo  {journal} {International Journal of Aeroacoustics}\ }\textbf {\bibinfo {volume} {18}},\ \bibinfo {pages} {317} (\bibinfo {year} {2019})}\BibitemShut {NoStop}%
\bibitem [{\citenamefont {Jaiswal}\ \emph {et~al.}(2022)\citenamefont {Jaiswal}, \citenamefont {Pasco}, \citenamefont {Yakhina},\ and\ \citenamefont {Moreau}}]{jaiswal2022experimental}%
  \BibitemOpen
  \bibfield  {author} {\bibinfo {author} {\bibfnamefont {P.}~\bibnamefont {Jaiswal}}, \bibinfo {author} {\bibfnamefont {Y.}~\bibnamefont {Pasco}}, \bibinfo {author} {\bibfnamefont {G.}~\bibnamefont {Yakhina}},\ and\ \bibinfo {author} {\bibfnamefont {S.}~\bibnamefont {Moreau}},\ }\bibfield  {title} {\bibinfo {title} {Experimental investigation of aerofoil tonal noise at low mach number},\ }\href@noop {} {\bibfield  {journal} {\bibinfo  {journal} {J. Fluid Mech.}\ }\textbf {\bibinfo {volume} {932}},\ \bibinfo {pages} {A37} (\bibinfo {year} {2022})}\BibitemShut {NoStop}%
\bibitem [{\citenamefont {Glegg}\ and\ \citenamefont {Devenport}(2017)}]{glegg2017aeroacoustics}%
  \BibitemOpen
  \bibfield  {author} {\bibinfo {author} {\bibfnamefont {S.}~\bibnamefont {Glegg}}\ and\ \bibinfo {author} {\bibfnamefont {W.}~\bibnamefont {Devenport}},\ }\href@noop {} {\emph {\bibinfo {title} {Aeroacoustics of low Mach number flows: fundamentals, analysis, and measurement}}}\ (\bibinfo  {publisher} {Academic Press},\ \bibinfo {year} {2017})\BibitemShut {NoStop}%
\bibitem [{\citenamefont {Roger}\ and\ \citenamefont {Moreau}(2004)}]{Roger2004}%
  \BibitemOpen
  \bibfield  {author} {\bibinfo {author} {\bibfnamefont {M.}~\bibnamefont {Roger}}\ and\ \bibinfo {author} {\bibfnamefont {S.}~\bibnamefont {Moreau}},\ }\bibfield  {title} {\bibinfo {title} {Broadband {S}elf {N}oise from {L}oaded {F}an {B}lades},\ }\href@noop {} {\bibfield  {journal} {\bibinfo  {journal} {AIAA J.}\ }\textbf {\bibinfo {volume} {42}},\ \bibinfo {pages} {536} (\bibinfo {year} {2004})}\BibitemShut {NoStop}%
\bibitem [{\citenamefont {Moreau}\ and\ \citenamefont {Roger}(2005)}]{Moreau2005}%
  \BibitemOpen
  \bibfield  {author} {\bibinfo {author} {\bibfnamefont {S.}~\bibnamefont {Moreau}}\ and\ \bibinfo {author} {\bibfnamefont {M.}~\bibnamefont {Roger}},\ }\bibfield  {title} {\bibinfo {title} {Effect of airfoil aerodynamic loading on trailing-edge noise sources},\ }\href@noop {} {\bibfield  {journal} {\bibinfo  {journal} {AIAA J.}\ }\textbf {\bibinfo {volume} {43}},\ \bibinfo {pages} {41} (\bibinfo {year} {2005})}\BibitemShut {NoStop}%
\bibitem [{\citenamefont {Wang}\ \emph {et~al.}(2009)\citenamefont {Wang}, \citenamefont {Moreau}, \citenamefont {Iaccarino},\ and\ \citenamefont {Roger}}]{Wang2009}%
  \BibitemOpen
  \bibfield  {author} {\bibinfo {author} {\bibfnamefont {M.}~\bibnamefont {Wang}}, \bibinfo {author} {\bibfnamefont {S.}~\bibnamefont {Moreau}}, \bibinfo {author} {\bibfnamefont {G.}~\bibnamefont {Iaccarino}},\ and\ \bibinfo {author} {\bibfnamefont {M.}~\bibnamefont {Roger}},\ }\bibfield  {title} {\bibinfo {title} {{LES} prediction of wall-pressure fluctuations and noise of a low-speed airfoil},\ }\href@noop {} {\bibfield  {journal} {\bibinfo  {journal} {Int. J. of Aeroacoustics}\ }\textbf {\bibinfo {volume} {8}},\ \bibinfo {pages} {177} (\bibinfo {year} {2009})}\BibitemShut {NoStop}%
\bibitem [{\citenamefont {Neal}(2010)}]{neal2010effects}%
  \BibitemOpen
  \bibfield  {author} {\bibinfo {author} {\bibfnamefont {D.~R.}\ \bibnamefont {Neal}},\ }\emph {\bibinfo {title} {The effects of rotation on the flow field over a controlled-diffusion airfoil}},\ \href@noop {} {Ph.D. thesis} (\bibinfo {year} {2010})\BibitemShut {NoStop}%
\bibitem [{\citenamefont {Lallier-Daniels}\ \emph {et~al.}(2013)\citenamefont {Lallier-Daniels}, \citenamefont {Moreau}, \citenamefont {Sanjos{\'e}},\ and\ \citenamefont {P{\'e}rot}}]{lallier2013numerical}%
  \BibitemOpen
  \bibfield  {author} {\bibinfo {author} {\bibfnamefont {D.}~\bibnamefont {Lallier-Daniels}}, \bibinfo {author} {\bibfnamefont {S.}~\bibnamefont {Moreau}}, \bibinfo {author} {\bibfnamefont {M.}~\bibnamefont {Sanjos{\'e}}},\ and\ \bibinfo {author} {\bibfnamefont {F.}~\bibnamefont {P{\'e}rot}},\ }\bibfield  {title} {\bibinfo {title} {Numerical analysis of axial fans for performance and noise evaluation using the lattice boltzmann method},\ }\href@noop {} {\bibfield  {journal} {\bibinfo  {journal} {CFD Canada}\ } (\bibinfo {year} {2013})}\BibitemShut {NoStop}%
\bibitem [{\citenamefont {Wu}\ \emph {et~al.}(2020)\citenamefont {Wu}, \citenamefont {Moreau},\ and\ \citenamefont {Sandberg}}]{Wu2020}%
  \BibitemOpen
  \bibfield  {author} {\bibinfo {author} {\bibfnamefont {H.}~\bibnamefont {Wu}}, \bibinfo {author} {\bibfnamefont {S.}~\bibnamefont {Moreau}},\ and\ \bibinfo {author} {\bibfnamefont {R.}~\bibnamefont {Sandberg}},\ }\bibfield  {title} {\bibinfo {title} {{On the noise generated by a controlled-diffusion aerofoil at $Re_c=1.5\times 10^5$}},\ }\href {https://doi.org/10.1017/jfm.2019.129} {\bibfield  {journal} {\bibinfo  {journal} {J.~Sound Vib.}\ }\textbf {\bibinfo {volume} {506}},\ \bibinfo {pages} {116152: 1} (\bibinfo {year} {2020})}\BibitemShut {NoStop}%
\bibitem [{\citenamefont {Jaiswal}\ \emph {et~al.}(2020)\citenamefont {Jaiswal}, \citenamefont {Moreau}, \citenamefont {Avallone}, \citenamefont {Ragni},\ and\ \citenamefont {Pr{\"o}bsting}}]{jaiswal2020use}%
  \BibitemOpen
  \bibfield  {author} {\bibinfo {author} {\bibfnamefont {P.}~\bibnamefont {Jaiswal}}, \bibinfo {author} {\bibfnamefont {S.}~\bibnamefont {Moreau}}, \bibinfo {author} {\bibfnamefont {F.}~\bibnamefont {Avallone}}, \bibinfo {author} {\bibfnamefont {D.}~\bibnamefont {Ragni}},\ and\ \bibinfo {author} {\bibfnamefont {S.}~\bibnamefont {Pr{\"o}bsting}},\ }\bibfield  {title} {\bibinfo {title} {On the use of two-point velocity correlation in wall-pressure models for turbulent flow past a trailing edge under adverse pressure gradient},\ }\href@noop {} {\bibfield  {journal} {\bibinfo  {journal} {Phys. of Fluids}\ }\textbf {\bibinfo {volume} {32}} (\bibinfo {year} {2020})}\BibitemShut {NoStop}%
\bibitem [{\citenamefont {Moreau}\ \emph {et~al.}(2016{\natexlab{b}})\citenamefont {Moreau}, \citenamefont {Laffay}, \citenamefont {Idier},\ and\ \citenamefont {Atalla}}]{Moreau2016}%
  \BibitemOpen
  \bibfield  {author} {\bibinfo {author} {\bibfnamefont {S.}~\bibnamefont {Moreau}}, \bibinfo {author} {\bibfnamefont {P.}~\bibnamefont {Laffay}}, \bibinfo {author} {\bibfnamefont {A.}~\bibnamefont {Idier}},\ and\ \bibinfo {author} {\bibfnamefont {N.}~\bibnamefont {Atalla}},\ }\bibfield  {title} {\bibinfo {title} {{Several noise controls of the trailing-edge noise of a Controlled-Diffusion airfoil}},\ }in\ \href@noop {} {\emph {\bibinfo {booktitle} {{22$^{\rm nd}$ AIAA/CEAS Aeroacoustics Conference}}}},\ \bibinfo {series and number} {\bibinfo {number} {AIAA 2016-2816 paper}}\ (\bibinfo {year} {2016})\BibitemShut {NoStop}%
\bibitem [{\citenamefont {Schlichting}\ and\ \citenamefont {Gersten}(2016)}]{schlichting2016boundary}%
  \BibitemOpen
  \bibfield  {author} {\bibinfo {author} {\bibfnamefont {H.}~\bibnamefont {Schlichting}}\ and\ \bibinfo {author} {\bibfnamefont {K.}~\bibnamefont {Gersten}},\ }\href@noop {} {\emph {\bibinfo {title} {Boundary-layer theory}}}\ (\bibinfo  {publisher} {springer},\ \bibinfo {year} {2016})\BibitemShut {NoStop}%
\bibitem [{\citenamefont {Zhu}\ \emph {et~al.}(2018)\citenamefont {Zhu}, \citenamefont {Lallier-Daniels}, \citenamefont {Sanjos\'e}, \citenamefont {Moreau},\ and\ \citenamefont {Carolus}}]{Zhu2018}%
  \BibitemOpen
  \bibfield  {author} {\bibinfo {author} {\bibfnamefont {T.}~\bibnamefont {Zhu}}, \bibinfo {author} {\bibfnamefont {D.}~\bibnamefont {Lallier-Daniels}}, \bibinfo {author} {\bibfnamefont {M.}~\bibnamefont {Sanjos\'e}}, \bibinfo {author} {\bibfnamefont {S.}~\bibnamefont {Moreau}},\ and\ \bibinfo {author} {\bibfnamefont {T.}~\bibnamefont {Carolus}},\ }\bibfield  {title} {\bibinfo {title} {{Rotating coherent flow structures as a source for narrow band tip clearance noise from axial fans}},\ }\bibfield  {booktitle} {\emph {\bibinfo {booktitle} {{22nd AIAA/CEAS Aeroacoustics Conference}}},\ }\href@noop {} {\bibfield  {journal} {\bibinfo  {journal} {J.~Sound Vib.}\ }\bibinfo {series} {{AIAA Paper 2016-2822}},\ \textbf {\bibinfo {volume} {418}},\ \bibinfo {pages} {198} (\bibinfo {year} {2018})}\BibitemShut {NoStop}%
\bibitem [{\citenamefont {Kr{\"o}mer}\ \emph {et~al.}(2019)\citenamefont {Kr{\"o}mer}, \citenamefont {Moreau},\ and\ \citenamefont {Becker}}]{Kroemer2019}%
  \BibitemOpen
  \bibfield  {author} {\bibinfo {author} {\bibfnamefont {F.}~\bibnamefont {Kr{\"o}mer}}, \bibinfo {author} {\bibfnamefont {S.}~\bibnamefont {Moreau}},\ and\ \bibinfo {author} {\bibfnamefont {S.}~\bibnamefont {Becker}},\ }\bibfield  {title} {\bibinfo {title} {{Experimental investigation of the interplay between the sound field and the flow field in skewed low-pressure axial fans}},\ }\href@noop {} {\bibfield  {journal} {\bibinfo  {journal} {J.~Sound Vib.}\ }\textbf {\bibinfo {volume} {442}},\ \bibinfo {pages} {220} (\bibinfo {year} {2019})}\BibitemShut {NoStop}%
\bibitem [{\citenamefont {Saraceno}\ \emph {et~al.}(2023)\citenamefont {Saraceno}, \citenamefont {Palleja~Cabre}, \citenamefont {Paruchuri},\ and\ \citenamefont {Ganapathisubramani}}]{saraceno2023influence}%
  \BibitemOpen
  \bibfield  {author} {\bibinfo {author} {\bibfnamefont {I.}~\bibnamefont {Saraceno}}, \bibinfo {author} {\bibfnamefont {S.}~\bibnamefont {Palleja~Cabre}}, \bibinfo {author} {\bibfnamefont {C.~C.}\ \bibnamefont {Paruchuri}},\ and\ \bibinfo {author} {\bibfnamefont {B.}~\bibnamefont {Ganapathisubramani}},\ }\bibfield  {title} {\bibinfo {title} {Influence of non-dimensional parameters on the tip leakage noise},\ }\href@noop {} {\bibfield  {journal} {\bibinfo  {journal} {29th AIAA/CEAS Aeroacoustics 2023 Conference, Paper number 3838}\ } (\bibinfo {year} {2023})}\BibitemShut {NoStop}%
\bibitem [{\citenamefont {You}\ \emph {et~al.}(2006)\citenamefont {You}, \citenamefont {Wang}, \citenamefont {Moin},\ and\ \citenamefont {Mittal}}]{you2006effects}%
  \BibitemOpen
  \bibfield  {author} {\bibinfo {author} {\bibfnamefont {D.}~\bibnamefont {You}}, \bibinfo {author} {\bibfnamefont {M.}~\bibnamefont {Wang}}, \bibinfo {author} {\bibfnamefont {P.}~\bibnamefont {Moin}},\ and\ \bibinfo {author} {\bibfnamefont {R.}~\bibnamefont {Mittal}},\ }\bibfield  {title} {\bibinfo {title} {Effects of tip-gap size on the tip-leakage flow in a turbomachinery cascade},\ }\href@noop {} {\bibfield  {journal} {\bibinfo  {journal} {Phys. of Fluids}\ }\textbf {\bibinfo {volume} {18}} (\bibinfo {year} {2006})}\BibitemShut {NoStop}%
\bibitem [{\citenamefont {Bendat}(1978)}]{bendat1978statistical}%
  \BibitemOpen
  \bibfield  {author} {\bibinfo {author} {\bibfnamefont {J.~S.}\ \bibnamefont {Bendat}},\ }\bibfield  {title} {\bibinfo {title} {Statistical errors in measurement of coherence functions and input/output quantities},\ }\href@noop {} {\bibfield  {journal} {\bibinfo  {journal} {J. of Sound and Vib.}\ }\textbf {\bibinfo {volume} {59}},\ \bibinfo {pages} {405} (\bibinfo {year} {1978})}\BibitemShut {NoStop}%
\bibitem [{\citenamefont {Roger}\ and\ \citenamefont {Moreau}(2005)}]{Roger2005}%
  \BibitemOpen
  \bibfield  {author} {\bibinfo {author} {\bibfnamefont {M.}~\bibnamefont {Roger}}\ and\ \bibinfo {author} {\bibfnamefont {S.}~\bibnamefont {Moreau}},\ }\bibfield  {title} {\bibinfo {title} {Back-scattering correction and further extensions of amiet's trailing edge noise model. part 1: theory},\ }\href@noop {} {\bibfield  {journal} {\bibinfo  {journal} {J.~Sound Vib.}\ }\textbf {\bibinfo {volume} {286}},\ \bibinfo {pages} {477} (\bibinfo {year} {2005})}\BibitemShut {NoStop}%
\bibitem [{\citenamefont {Moreau}\ and\ \citenamefont {Roger}(2009)}]{Moreau2009}%
  \BibitemOpen
  \bibfield  {author} {\bibinfo {author} {\bibfnamefont {S.}~\bibnamefont {Moreau}}\ and\ \bibinfo {author} {\bibfnamefont {M.}~\bibnamefont {Roger}},\ }\bibfield  {title} {\bibinfo {title} {{Back-scattering correction and further extensions of Amiet's trailing-edge noise model. Part II: Application}},\ }\href {https://doi.org/10.1016/j.jsv.2008.11.051} {\bibfield  {journal} {\bibinfo  {journal} {J.~Sound Vib.}\ }\textbf {\bibinfo {volume} {323}},\ \bibinfo {pages} {397} (\bibinfo {year} {2009})}\BibitemShut {NoStop}%
\bibitem [{\citenamefont {Moreau}\ \emph {et~al.}(2007)\citenamefont {Moreau}, \citenamefont {Schram},\ and\ \citenamefont {Roger}}]{Moreau2007}%
  \BibitemOpen
  \bibfield  {author} {\bibinfo {author} {\bibfnamefont {S.}~\bibnamefont {Moreau}}, \bibinfo {author} {\bibfnamefont {C.}~\bibnamefont {Schram}},\ and\ \bibinfo {author} {\bibfnamefont {M.}~\bibnamefont {Roger}},\ }\bibfield  {title} {\bibinfo {title} {Diffraction effects on the trailing edge noise measured in an open-jet anechoic wind tunnel},\ }in\ \href@noop {} {\emph {\bibinfo {booktitle} {13th AIAA/CEAS Aeroacoustics Conference Meeting and Exhibit, Monterey, CA}}},\ \bibinfo {series and number} {AIAA 2007-3706 paper}\ (\bibinfo {year} {2007})\BibitemShut {NoStop}%
\bibitem [{\citenamefont {Henning}\ \emph {et~al.}(2008)\citenamefont {Henning}, \citenamefont {Kaepernick}, \citenamefont {Ehrenfried}, \citenamefont {Koop},\ and\ \citenamefont {Dillmann}}]{henning2008investigation}%
  \BibitemOpen
  \bibfield  {author} {\bibinfo {author} {\bibfnamefont {A.}~\bibnamefont {Henning}}, \bibinfo {author} {\bibfnamefont {K.}~\bibnamefont {Kaepernick}}, \bibinfo {author} {\bibfnamefont {K.}~\bibnamefont {Ehrenfried}}, \bibinfo {author} {\bibfnamefont {L.}~\bibnamefont {Koop}},\ and\ \bibinfo {author} {\bibfnamefont {A.}~\bibnamefont {Dillmann}},\ }\bibfield  {title} {\bibinfo {title} {Investigation of aeroacoustic noise generation by simultaneous particle image velocimetry and microphone measurements},\ }\href@noop {} {\bibfield  {journal} {\bibinfo  {journal} {Exp. in fluids}\ }\textbf {\bibinfo {volume} {45}},\ \bibinfo {pages} {1073} (\bibinfo {year} {2008})}\BibitemShut {NoStop}%
\bibitem [{\citenamefont {Brooks}\ and\ \citenamefont {Hodgson}(1981)}]{brooks1981trailing}%
  \BibitemOpen
  \bibfield  {author} {\bibinfo {author} {\bibfnamefont {T.~F.}\ \bibnamefont {Brooks}}\ and\ \bibinfo {author} {\bibfnamefont {T.}~\bibnamefont {Hodgson}},\ }\bibfield  {title} {\bibinfo {title} {Trailing edge noise prediction from measured surface pressures},\ }\href@noop {} {\bibfield  {journal} {\bibinfo  {journal} {Journal of sound and vibration}\ }\textbf {\bibinfo {volume} {78}},\ \bibinfo {pages} {69} (\bibinfo {year} {1981})}\BibitemShut {NoStop}%
\bibitem [{\citenamefont {Yu}\ and\ \citenamefont {Tam}(1978)}]{yu1978experimental}%
  \BibitemOpen
  \bibfield  {author} {\bibinfo {author} {\bibfnamefont {J.}~\bibnamefont {Yu}}\ and\ \bibinfo {author} {\bibfnamefont {C.}~\bibnamefont {Tam}},\ }\bibfield  {title} {\bibinfo {title} {Experimental investigation of the trailing edge noise mechanism},\ }\href@noop {} {\bibfield  {journal} {\bibinfo  {journal} {AIAA J.}\ }\textbf {\bibinfo {volume} {16}},\ \bibinfo {pages} {1046} (\bibinfo {year} {1978})}\BibitemShut {NoStop}%
\bibitem [{\citenamefont {Williams}\ and\ \citenamefont {Hall}(1970)}]{williams1970aerodynamic}%
  \BibitemOpen
  \bibfield  {author} {\bibinfo {author} {\bibfnamefont {J.~F.}\ \bibnamefont {Williams}}\ and\ \bibinfo {author} {\bibfnamefont {L.}~\bibnamefont {Hall}},\ }\bibfield  {title} {\bibinfo {title} {Aerodynamic sound generation by turbulent flow in the vicinity of a scattering half plane},\ }\href@noop {} {\bibfield  {journal} {\bibinfo  {journal} {Journal of fluid mechanics}\ }\textbf {\bibinfo {volume} {40}},\ \bibinfo {pages} {657} (\bibinfo {year} {1970})}\BibitemShut {NoStop}%
\bibitem [{\citenamefont {Moreau}\ and\ \citenamefont {Doolan}(2016)}]{moreau2016experimental}%
  \BibitemOpen
  \bibfield  {author} {\bibinfo {author} {\bibfnamefont {D.~J.}\ \bibnamefont {Moreau}}\ and\ \bibinfo {author} {\bibfnamefont {C.~J.}\ \bibnamefont {Doolan}},\ }\bibfield  {title} {\bibinfo {title} {An experimental study of airfoil tip vortex formation noise},\ }in\ \href@noop {} {\emph {\bibinfo {booktitle} {Proceedings of acoustics}}},\ Vol.~\bibinfo {volume} {2}\ (\bibinfo {year} {2016})\ pp.\ \bibinfo {pages} {1167--1176}\BibitemShut {NoStop}%
\bibitem [{\citenamefont {Sirovich}(1987)}]{sirovich1987turbulence}%
  \BibitemOpen
  \bibfield  {author} {\bibinfo {author} {\bibfnamefont {L.}~\bibnamefont {Sirovich}},\ }\bibfield  {title} {\bibinfo {title} {Turbulence and the dynamics of coherent structures. i. coherent structures},\ }\href@noop {} {\bibfield  {journal} {\bibinfo  {journal} {Quarterly of applied mathematics}\ }\textbf {\bibinfo {volume} {45}},\ \bibinfo {pages} {561} (\bibinfo {year} {1987})}\BibitemShut {NoStop}%
\bibitem [{\citenamefont {Koch}(2021)}]{Koch_diss}%
  \BibitemOpen
  \bibfield  {author} {\bibinfo {author} {\bibfnamefont {R.}~\bibnamefont {Koch}},\ }\emph {\bibinfo {title} {{Identification des sources de bruit aérodynamique liées aux écoulements de jeu en tête de pale de soufflante de turboréacteur}}},\ \href@noop {} {Ph.D. thesis},\ \bibinfo  {school} {University of Sherbrooke}, \bibinfo {address} {"https://savoirs.usherbrooke.ca/handle/11143/18703"} (\bibinfo {year} {2021})\BibitemShut {NoStop}%
\bibitem [{\citenamefont {Pearl}\ \emph {et~al.}(2000)\citenamefont {Pearl} \emph {et~al.}}]{pearl2000models}%
  \BibitemOpen
  \bibfield  {author} {\bibinfo {author} {\bibfnamefont {J.}~\bibnamefont {Pearl}} \emph {et~al.},\ }\bibinfo {title} {Models, reasoning and inference}\ (\bibinfo  {publisher} {Cambridge University Press},\ \bibinfo {address} {Cambridge, UK},\ \bibinfo {year} {2000})\ p.\ \bibinfo {pages} {386}\BibitemShut {NoStop}%
\bibitem [{\citenamefont {Chung}\ \emph {et~al.}(2018)\citenamefont {Chung}, \citenamefont {Monty},\ and\ \citenamefont {Hutchins}}]{chung2018similarity}%
  \BibitemOpen
  \bibfield  {author} {\bibinfo {author} {\bibfnamefont {D.}~\bibnamefont {Chung}}, \bibinfo {author} {\bibfnamefont {J.~P.}\ \bibnamefont {Monty}},\ and\ \bibinfo {author} {\bibfnamefont {N.}~\bibnamefont {Hutchins}},\ }\bibfield  {title} {\bibinfo {title} {Similarity and structure of wall turbulence with lateral wall shear stress variations},\ }\href@noop {} {\bibfield  {journal} {\bibinfo  {journal} {J. Fluid Mech.}\ }\textbf {\bibinfo {volume} {847}},\ \bibinfo {pages} {591} (\bibinfo {year} {2018})}\BibitemShut {NoStop}%
\bibitem [{\citenamefont {Palleja-Cabre}\ \emph {et~al.}(2022)\citenamefont {Palleja-Cabre}, \citenamefont {Saraceno}, \citenamefont {Paruchuri},\ and\ \citenamefont {Joseph}}]{palleja2022reduction}%
  \BibitemOpen
  \bibfield  {author} {\bibinfo {author} {\bibfnamefont {S.}~\bibnamefont {Palleja-Cabre}}, \bibinfo {author} {\bibfnamefont {I.}~\bibnamefont {Saraceno}}, \bibinfo {author} {\bibfnamefont {C.~C.}\ \bibnamefont {Paruchuri}},\ and\ \bibinfo {author} {\bibfnamefont {P.}~\bibnamefont {Joseph}},\ }\bibfield  {title} {\bibinfo {title} {Reduction of tip-leakage noise by using porosity},\ }\href@noop {} {\bibfield  {journal} {\bibinfo  {journal} {28th AIAA/CEAS Aeroacoustics 2022 Conference, Paper number 3061}\ } (\bibinfo {year} {2022})}\BibitemShut {NoStop}%
\bibitem [{\citenamefont {Jaiswal}\ and\ \citenamefont {Ganapathisubramani}(2024)}]{jaiswal2023effects}%
  \BibitemOpen
  \bibfield  {author} {\bibinfo {author} {\bibfnamefont {P.}~\bibnamefont {Jaiswal}}\ and\ \bibinfo {author} {\bibfnamefont {B.}~\bibnamefont {Ganapathisubramani}},\ }\bibfield  {title} {\bibinfo {title} {Effects of porous substrates on the structure of turbulent boundary layers},\ }\href@noop {} {\bibfield  {journal} {\bibinfo  {journal} {J. Fluid Mech.}\ }\textbf {\bibinfo {volume} {980}},\ \bibinfo {pages} {A39} (\bibinfo {year} {2024})}\BibitemShut {NoStop}%
\end{thebibliography}%

\end{document}